\documentclass[letterpaper,twocolumn,10pt]{article}
\usepackage{usenix}
\usepackage{epsfig}
\usepackage{comment}
\usepackage[pdftex]{color}
\usepackage{url}
\usepackage{listings}
\usepackage{textcomp}
\usepackage{gnuplot-lua-tikz}
\usepackage{amssymb}
\usepackage{amsmath}
\usepackage{paralist}
\usepackage{subcaption}
\usepackage{algorithm}
\usepackage{algpseudocode}
\usepackage{graphicx}
\definecolor{lbcolor}{rgb}{0.9,0.9,0.9}

\newlength{\saveparindent}
\newlength{\saveparskip}
\begin{document}

\date{}

\title{An SSD-based eigensolver for spectral analysis on billion-node graphs}

\author{
{\rm Da Zheng, Randal Burns}\\
Department of Computer Science \\
Johns Hopkins University
\and
{\rm Joshua Vogelstein}\\
Institute for Computational Medicine \\
Department of Biomedical Engineering \\
Johns Hopkins University
\and
{\rm Carey E. Priebe}\\
Department of Applied Mathematics and Statistics \\
Johns Hopkins University
\and
{\rm Alexander S. Szalay}\\
Department of Physics and Astronomy \\
Johns Hopkins University
} 

\maketitle

\thispagestyle{empty}

\subsection*{Abstract}
Many eigensolvers such as ARPACK and Anasazi have been developed to compute
eigenvalues of a large sparse matrix. These eigensolvers are limited by
the capacity of RAM. They run in memory of a single machine for smaller
eigenvalue problems and require the distributed memory for larger problems.

In contrast, we develop an SSD-based eigensolver framework called FlashEigen,
which extends Anasazi eigensolvers to SSDs, to compute eigenvalues of a graph
with hundreds of millions or even billions of vertices in a single machine.
FlashEigen performs sparse matrix multiplication in a semi-external memory
fashion, i.e., we keep the sparse matrix on SSDs and the dense matrix in memory.
We store the entire vector subspace on SSDs and reduce I/O to improve
performance through caching the most recent dense matrix.
Our result shows that FlashEigen is able to achieve 40\%-60\% performance
of its in-memory implementation and has performance comparable to the Anasazi
eigensolvers on a machine with 48 CPU cores. Furthermore, it is capable of
scaling to a graph with 3.4 billion vertices and 129 billion edges. It takes
about four hours to compute eight eigenvalues of the billion-node graph using
120 GB memory.

\section{Introduction}
Spectral analysis \cite{} is a fundamental tool for both graph analysis and
other areas of data mining. Essentially, it computes
eigenvalues and eigenvectors of graphs to infer properties of graphs.
spectral clustering \cite{Ng01, Sussman12}, triangle counting \cite{Tsourakakis08}.
Many real-world graphs are massive: Facebook's social network has billions
of vertices and today's web graph is even much larger.


It is computationally expensive to compute all eigenvalues and
eigenvectors of a large matrix. The computation complexity is
$O(n^3)$ on a square matrix \cite{Pan99}, where $n$ is the number of rows
and columns of the matrix.
When the size of a matrix grows to millions or even billions of rows and
columns, it becomes prohibitive to compute all eigenvalues and
eigenvectors.

Numerous algorithms \cite{Lanczos, IRLM, krylovschur, Arbenz05} have been
developed to compute a small number of eigenpairs.
The current popular eigensolver packages such as ARPACK \cite{arpack}
and Anasazi \cite{anasazi} have state-of-art eigensolvers
capable of computing a few eigenvalues with certain properties such as
the eigenvalues of the largest or smallest magnitude. All of these eigensolvers
perform a sequence of sparse matrix multiplication to construct and update
a vector subspace $S \in \mathbb{R}^{n \times m}$, where $n$ is the size of
the eigenproblem and $m$ is the subspace size \cite{Arbenz05}. In addition,
they perform matrix operations on the vector subspace. When computing eigenpairs
of a graph at the billion scale, neither the sparse matrix that represents
a graph nor the vector subspace fits in the RAM of a single machine.

It is challenging to implement an efficient kernel of sparse matrix
multiplication for many real-world graphs. Sparse matrix multiplication
on these graphs induces many small random memory access due to near-random
vertex connection. It may suffer from load imbalancing because of
the power-law distribution in vertex degree. Furthermore, graphs cannot be
clustered or partitioned effectively \cite{leskovec} to localize access.

Large-scale eigendecomposition is generally solved in a large cluster
\cite{anasazi, slepc}, where the aggregate memory is sufficient to store
the sparse matrix and the vector subspace. Sparse matrix multiplication
on graphs in distributed memory leads to significant network communication
and is usually bottlenecked by the network. As such, this operation requires
fast network to achieve performance. However, a supercomputer or a large
cluster with fast network communication is not accessible to many people.


We build FlashEigen, an external-memory eigensolver, on top of a user-space
filesystem called SAFS \cite{safs} to solve a large eigenproblem with SSDs
in a single machine. 
Instead of developing a new eigensolver from scratch, we leverage
the Anasazi framework and implement SSD-based matrix operations
for the framework. FlashEigen is specifically optimized for the Block
Krylov-Schur \cite{krylovschur} eigensolver because it is the fastest and
generates the least I/O among the Anasazi eigensolvers when computing
eigenvalues of many power-law graphs. The Anasazi eigensolvers implement
a block extension, which enables the eigensolvers to update multiple vectors
in the subspace in each iteration to increase computation density and amortize
I/O overhead. As a result, the eigensolvers require sparse matrix dense matrix
multiplication (SpMM). Given a graph with hundreds of millions of vertices or
even billions of vertices,
the vector subspace constructed by the eigensolvers requires very large storage
size, usually much larger than the sparse matrix. Therefore, FlashEigen stores
the entire vector subspace on SSDs.


Although SSDs can deliver high IOPS and sequential I/O throughput, we have
to overcome many technical challenges to construct an external-memory
eigensolver with performance comparable to in-memory eigensolvers.
First, it is challenging to achieve the maximal I/O throughput, on the order
of ten gigabytes, from a large array of commodity SSDs, due to many overheads
from the operating system. Even achieving the maximal I/O throughput, SSDs are
still an order of magnitude slower than DRAM in throughput. Furthermore,
SSDs wear out after we write data to them. For example, some enterprise
SSDs \cite{ocz} only allows one DWPD (diskful writes per day). Writing too much
data to these SSDs drastically shortens their lives and increases operation
cost.

We perform SpMM in a semi-external memory (SEM) fashion, which keeps
the sparse matrix on SSDs and dense matrices in memory. This operation streams
rows in the sparse matrix to memory and multiplies with the dense matrix in
memory, which generates sequential I/O and allows us to yield maximal I/O
throughput from SSDs. While maximizing the I/O throughput, we also compress
the sparse matrix to further accelerate retrieving the sparse matrix from SSDs.
The SEM strategy incorporates well with in-memory optimizations for SpMM.
We deploy multiple in-memory optimizations specifically designed for power-law
graphs. For example, we assign partitions of the sparse matrix dynamically to
threads for load balancing, deploy cache blocking to increase CPU cache hits,
partition and evenly distribute the dense matrix to NUMA nodes to fully utilize
the memory bandwidth of a NUMA machine.

We deploy optimizations on dense matrix operations to reduce I/O and fully
utilize the I/O bandwidth of the SSDs. Thanks to the block extension of
the Anasazi eigensolvers, we group multiple vectors together into a column-major
dense matrix and store each dense matrix in a separate SAFS file for efficient
I/O access to any vectors in the subspace. When accessing a large number of dense
matrices in a single operation, we group dense matrices to constrain memory
consumption of the matrix operation to improve its scalability. To reduce I/O
and alleviate SSD wear out, we use deploy lazy evaluation and cache the most
recent dense matrix in the subspace.


Our result shows that for many real-world sparse graphs, the SSD-based
eigensolver is able to achieve 40\%-60\%
performance of its in-memory implementation and has performance comparable to
the Anasazi eigensolver on a machine with 48 CPU cores for computing various
numbers of eigenvalues. We further demonstrate that the SSD-based eigensolver
is capable of scaling to a graph with 3.4 billion vertices and 129 billion edges.
It takes about 4 hours to compute eight eigenvalues of the billion-node graph
and use 120 GB memory. We conclude that our solution offers new design
possibilities for large-scale eigendecomposition, replacing memory with larger
and cheaper SSDs and processing bigger problems on fewer machines.

\section{Related Work}
Anasazi \cite{anasazi} is an eigensolver framework in the Trilinos project
\cite{trilinos}. This framework implements block extension of multiple
eigensolver algorithms
such as Block Krylov-Schur method \cite{krylovschur}, Block Davidson method
\cite{Arbenz05} and LOBPCG \cite{Arbenz05}. This is a very flexible framework
that allows users to redefine sparse matrix dense matrix multiplication and
dense matrix operations. By default, Anasazi uses the matrix implementations
in Trilinos that run in the distributed memory.

Arpack \cite{arpack} is another state-of-art eigensolver commonly used by
multiple numeric computation frameworks such as Matlab. This eigensolver
implements the implicitly restarted arnoldi method \cite{IRAM}. Arpack
only allows users to redefine sparse matrix vector multiplication.
Its dense matrix operations by default run in serial.

Sparse matrix vector multiplication (SpMV) and sparse matrix dense matrix
multiplication (SpMM) are an important operation in numeric computation and
are well studied in the literature. For example, Williams et. al
\cite{Williams07} described multiple optimizations for sparse matrix
vector multiplication in multicore architecture. Yoo et. al \cite{Yoo11}
and Boman et. al \cite{Boman2013} optimized SpMV for large scale-free graphs
using 2D graph partitioning. Aktulga et. al \cite{Aktulga14} optimized sparse
matrix dense matrix multiplication with cache blocking. In contrast, we
further advance sparse matrix dense matrix multiplication with a focus on
optimizations for external memory.

FlashGraph \cite{flashgraph} is a general graph analysis framework. It performs
graph algorithms in a semi-external memory fashion \cite{sem}, i.e., it keeps
vertex state in memory and edge lists on SSDs. It is specifically optimized for
graph algorithms that has a fraction of vertices running in each iteration.
This design prevents FlashGraph from performing some optimizations for sparse
matrix multiplication as shown in this paper.

HEIGEN \cite{Kang11} is an eigensolver implemented with MapReduce \cite{mapreduce}
to compute eigenpairs for spectral graph analysis. HEIGEN implements a basic
Lanczos algorithm
\cite{Lanczos} with selective orthogonalization from scratch. In contrast, our
approach extends the state-of-art implementations to SSDs. By integrating many
SSDs to a single machine, our approach can compete with a cluster.

Zhou et al. \cite{Zhou12} implemented an LOBPCG \cite{Arbenz05} eigensolver in
an SSD cluster. Their implementation targets nuclear many-body Hamiltonian
matrices, which are much denser and have smaller dimensions than many sparse
graphs. Therefore, their solution stores the sparse matrix on SSDs and keep
the entire vector subspace in RAM. Their solution also focus on optimizations
in the distributed environment. In contrast, we store both the sparse matrix
and the vector subspace on SSDs due to the large number of vertices in
our target graphs. We focus on external-memory optimizations in a single machine.

\section{Design}
\begin{figure}
\centering
\includegraphics[scale=0.4]{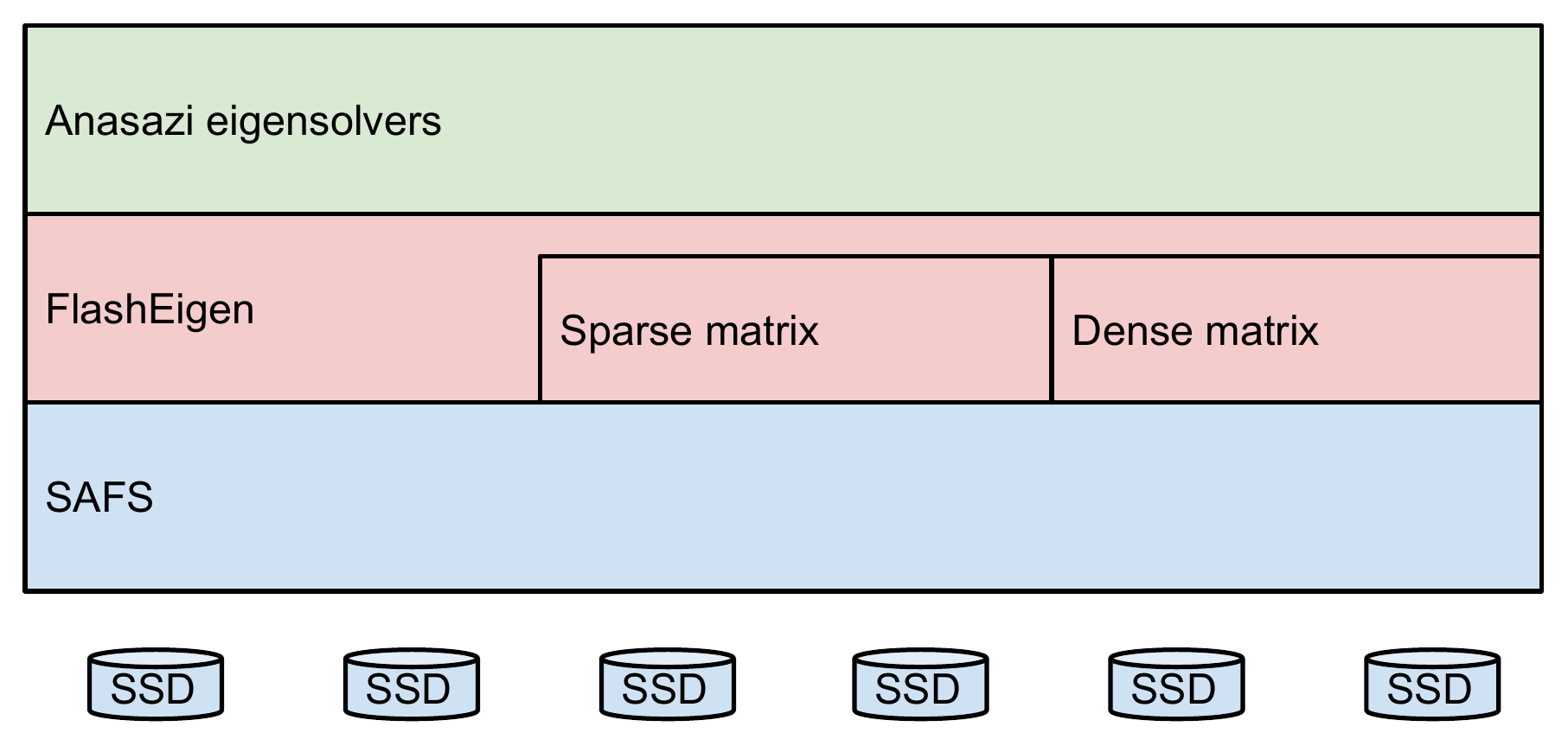}
\vspace{-5pt}
\caption{The architecture of FlashEigen.}
\vspace{-5pt}
\label{arch}
\end{figure}

FlashEigen is an external-memory eigensolver framework optimized for any fast
I/O devices
such as a large SSD array to compute eigenvalues of sparse graphs (Figure
\ref{arch}). We build FlashEigen on top of SAFS, a user-space filesystem,
to fully utilize the I/O throughput of a large SSD array. Instead of
implementing eigenvalue algorithms from scratch, we integrate FlashEigen
with the Anasazi framework to compute eigenvalues. The Anasazi framework
implements multiple state-of-art eigenvalue algorithms and provides users
a flexible programming interface to redefine both sparse and dense matrix
operations required by the eigenvalue algorithms. FlashEigen stores both sparse
and dense matrices in SAFS and focuses on optimizing the matrix operations
required by the Anasazi eigensolvers for SSDs.

\subsection{Eigensolver algorithm}


\begin{algorithm}
	\begin{algorithmic}[1]
		\For{i = 0, 1, ..., until convergence}
		\State (1) Update the subspace $S \in \mathbb{R}^{n \times m}$,
		\State (2) Solve the projected eigenproblem $S^TASy = S^TSy\theta$.
		\State (3) Compute the residual: $r = Kx - x\theta$, where
		\State\hspace{\algorithmicindent} $x = Sy$ (Ritz vector), $\theta = \rho(x)$ (Ritz value).
		\State (4) Test the convergence of a Ritz pair $(x, \rho(x))$.
		\EndFor
	\end{algorithmic}
	\caption{Pseudo code of a generic eigenvalue algorithm that compute eigenvalues
	of a square matrix $A$ with $n$ rows and columns.}
	\label{eigencode}
\end{algorithm}

The state-of-art eigenvalue algorithms compute eigenvalues with iterative
methods. Algorithm \ref{eigencode} shows the general steps used by the eigenvalue
algorithms.
Step (1) constructs a vector subspace $S \in \mathbb{R}^{n \times m}$, where
$n$ is the number of rows and columns of a sparse matrix and $m$ is the number
of vectors in the subspace. When computing eigenvalues of a sparse graph,
two key operations in this step are sparse matrix multiplication to construct
the subspace and reorthogonalization to correct floating-point rounding errors.
Step (2) projects the large sparse matrix to a much smaller matrix with only
$m$ rows and $m$ columns, which can be solved by other eigensolvers such as
the one in LAPACK \cite{lapack}. Step (3) projects the solution of the small
eigenvalue problem back to the original eigenvalue problem. Step (4) tests
whether the projected solution fulfills the precision requirement given by
users. If not, the algorithm adjusts the vector subspace, returns to step (1)
and continues the process.

The block extenion implemented in the Anasazi eigensolvers updates multiple
vectors in the subspace in an iteration. This optimization leads to sparse
matrix dense matrix multiplication, which increases computation density to
improve performance, as well as a set of dense matrix operations instead of
vector operations. The number of vectors to be updated together is determined
by the \textit{block size}, denoted by $b$. The subspace size $m$
is $b \times NB$, where $NB$ is the number of blocks.

\subsection{SAFS}

SAFS \cite{safs} is a user-space filesystem for a high-speed SSD array in
a NUMA (non-uniform memory architecture) machine. It is implemented as
a library and runs in the address space
of its application. It is deployed on top of the Linux native filesystem.
SAFS was originally designed for optimizing small I/O accesses. However,
sparse matrix multiplication and dense matrix operations
generate much fewer but much larger I/O. Therefore, we provide additional
optimizations to maximize sequential I/O throughput from a large SSD array.

The latency of a thread context switch becomes noticeable on a high-speed SSD
array under a sequential I/O workload and it becomes critical to avoid thread
context switch to gain I/O performance. If the computation in application
threads does not saturate CPU, SAFS will put the application threads into
sleep while they are waiting for I/O. This results in many thread context
switches and underutilization of both CPU and SSDs. To saturate I/O,
an application thread issues asynchronous I/O and poll for I/O to avoid thread
context switches after completing all computation available to it.

To better support access to many relatively small files simultaneously, SAFS
stripes data in a file across SSDs with a different striping order for each file.
This strategy stores data from multiple files evenly across SSDs and improves
I/O utialization. Due to the sequential I/O workload, FlashEigen stripes data
across SSDs with a large block size, on the order of megabytes, to increase I/O
throughput and potentially reduce write amplification on SSDs \cite{Tang15}.
Such a large block size may cause storage skew for small files
on a large SSD array if every file stripes data in the same order. Using
the same striping order for all files may also cause skew in I/O access.
Therefore, SAFS generates a random striping order for each file to evenly
distribute I/O among SSDs when a file is created. SAFS stores the striping
order with the file for future data retrieval.

\subsection{Sparse matrix multiplication} \label{spmm}
Sparse matrix multiplication is one of the most computationally expensive
operations in an eigensolver. Applying this operation on real-world graphs
usually leads to many random memory accesses and its performance is usually
limited by the random memory performance of DRAM. The block extension
in the Anasazi eigensolvers enables sparse matrix dense matrix multiplication
to increase data locality and computation density and improve overall performance
of an eigensolver.

To scale sparse matrix multiplication to a sparse graph with billions of vertices,
we perform this operation in semi-external memory (SEM). That is, we keep dense
matrices in memory and the sparse
matrix on SSDs. This strategy enables nearly in-memory performance while achieving
the scalability in proportion to the ratio of edges to vertices in a graph.

\subsubsection{The sparse matrix format}
The state-of-art numeric libraries store a sparse matrix in compressed row storage
(CSR) or compressed column storage (CSC) format. However, these formats incur
many CPU cache misses in sparse matrix multiplication on many real-world graphs
due to their nearly random vertex connection. They also require a relatively
large storage size. For a graph with billions of edges, we have to use eight
bytes to store the row and column indices. For semi-external memory sparse
matrix multiplication, SSDs may become the bottleneck if a sparse matrix has
a large storage size.
Therefore, we need to use an alternative format for sparse matrices to increase
CPU cache hits and reduce the amount of data read from SSDs.

\begin{figure}
\centering
\includegraphics[scale=0.3]{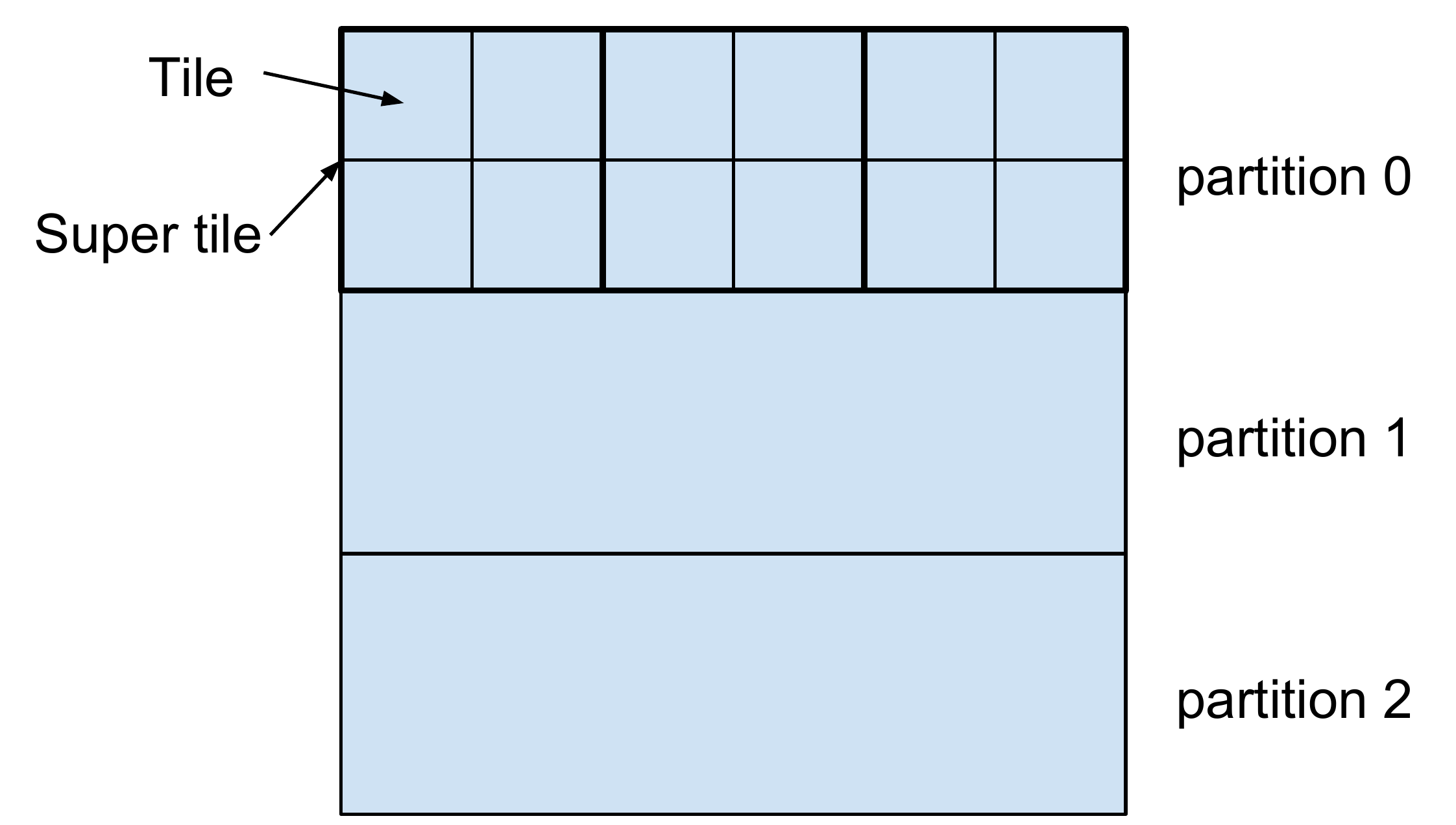}
\vspace{-5pt}
\caption{The format of a sparse matrix in FlashEigen.}
\vspace{-5pt}
\label{sparse_mat}
\end{figure}

To increase CPU cache hits, we deploy cache blocking \cite{Im04} and store
non-zero entries of a sparse matrix in tiles (Figure \ref{sparse_mat}).
When a tile is small, the rows from the input and output dense matrices
involved in the multiplication with the tile are always kept in the CPU cache
during the multiplication. The optimal tile size should fill the CPU cache
with the rows of the dense matrices involved in the multiplication with
the tile and is affected by the number of columns of the dense matrices,
which is chosen by users. Instead of generating a sparse matrix with
different tile sizes optimized for different numbers of columns in the dense
matrices, we use a relatively small tile size and rely on the runtime system
to optimize for different numbers of columns (in section \ref{sec:exec}).
In the semi-external memory, we expect that the dense matrices do not
have more than eight columns in sparse matrix multiplication. Therefore, we
use the tile size of $16K \times 16K$ by default to balance the matrix storage
size and the adaptibility to different numbers of columns.

\begin{figure}
\centering
\includegraphics[scale=0.5]{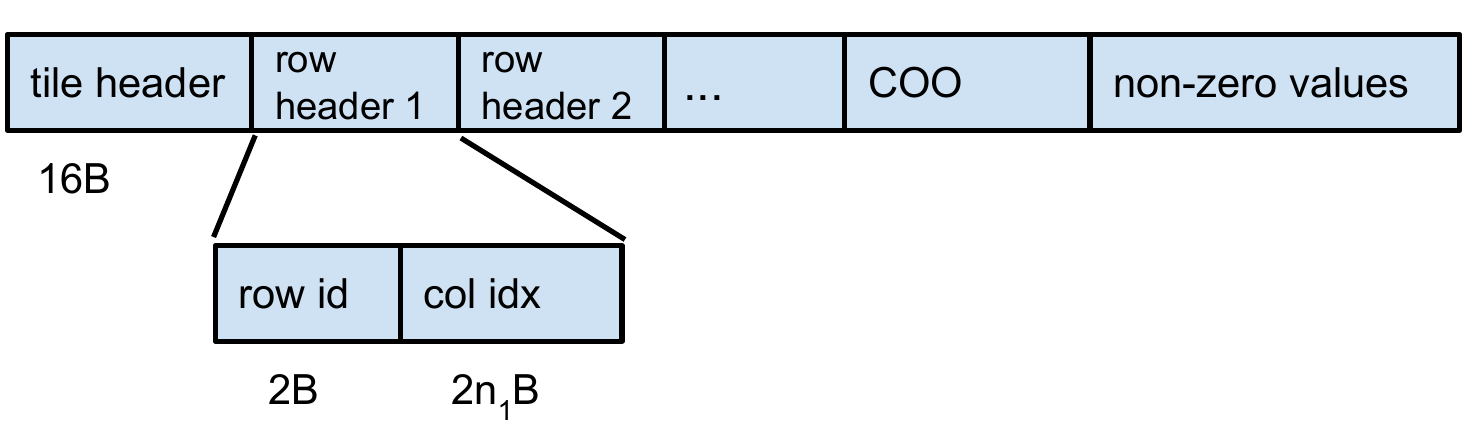}
\vspace{-5pt}
\caption{The storage format of a tile in a sparse matrix.}
\vspace{-5pt}
\label{tile_format}
\end{figure}

To reduce the overall storage size of a sparse matrix, we use a compact format
to store non-zero entries in a tile. In very sparse matrices such as
many real-world graphs, many rows in a tile do not have any non-zero entries.
The CSR (CSC) format requires an entry for each row (column) in the row
(column) index. Therefore, the CSR or CSC format wastes space when storing elements
in a tile. Instead, we only keep data for rows with non-zero entries in a tile
shown in Figure \ref{tile_format} and refer to this format as SCSR (Super
Compressed Row Storage). This format maintains a row header for each non-empty
row. A row header has an identifier to indicate the row number, followed by
column indices. 
The most significant bit of the identifier is always set to 1, while the most
significant bit of a column index entry is always set to 0. As such, we can easily
distinguish a row identifier from a column index entry and determine the end
of a row. Thanks to the small size of a tile, we use two bytes to further store a row
number and a column index entry to reduce the storage size. Since the most
significant bit is used to indicate the beginning of a row, this format allows
a maximum tile size of $32K \times 32K$.

For many real-world graphs, many rows in a tile have only one non-zero entry,
thanks to the sparsity of the graphs and nearly random vertex connection.
Iterating over single-entry rows requires to test the end of a row for every
non-zero entry, resulting in many extra conditional jump CPU instructions
in sparse matrix multiplication.
In contrast, the coordinate format (COO) is more suitable for storing these
single-entry rows. It does not increase the storage size but significantly
reduces the number of conditional jump instructions when we iterate
them. As a result, we hybrid SCSR and COO to store non-zero entries in a tile
with COO stored behind the row headers of SCSR. All non-zero entries are
stored together at the end of a tile.

We organize tiles in a sparse matrix in tile rows and maintain a matrix index
for them. Each entry of the index stores the location of a tile row on SSDs
to facilitate random access
to tile rows. This is useful for parallelizing sparse matrix multiplication.
Because a tile contains thousands of rows, the matrix index requires a very
small storage size even for a billion-node graph. We keep the entire index
in memory during sparse matrix multiplication.

\subsubsection{The dense matrix format for SpMM} \label{numa_mat}
Dense matrices in sparse matrix multiplication are tall-and-skinny (TAS)
matrices
with millions or even billions of rows but only several columns. The number
of rows in a dense matrix is determined by the number of vertices in a sparse
graph and the number of columns is determined by the \textit{block size}
in an eigensolver with the block extension. The dense matrix is kept in memory
for semi-external memory (SEM) sparse matrix dense matrix multiplication (SpMM),
so the size of the dense matrix governs the memory consumption
of SpMM. Given the limited amount of RAM in a machine, the number of columns
in a dense matrix has to be small.

\begin{figure}
\centering
\includegraphics[scale=0.4]{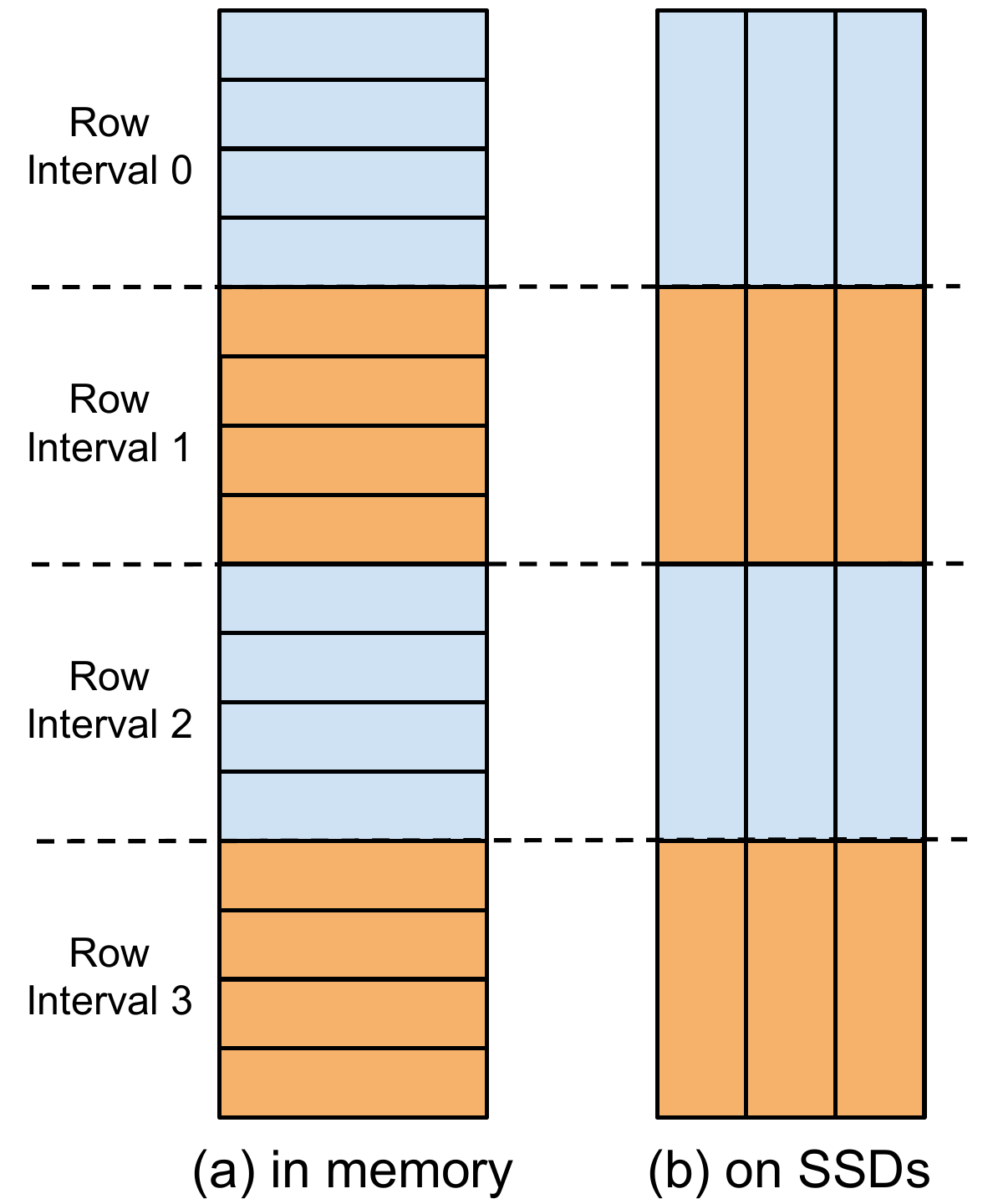}
\vspace{-5pt}
\caption{The data layout of tall-and-skinny (TAS) dense matrices. A TAS
dense matrix is partitioned horizontally into many row intervals.
(a) For an in-memory matrix, row intervals are stored across NUMA nodes and
elements are stored in row-major order; (b) for an SSD-based matrix, elements
inside a row interval are stored in column-major order.}
\vspace{-5pt}
\label{dense_mat}
\end{figure}

For a non-uniform memory architecture (NUMA), we partition the input dense matrix
horizontally and store partitions evenly across NUMA nodes to fully utilize
the bandwidth of memory and inter-processor links in sparse matrix
multiplication. The NUMA architecture is prevalent in today's multi-processor
servers, where each processor connects to its own memory banks. As shown in
Figure \ref{dense_mat} (a), we assign multiple
contiguous rows in a row interval to a partition, which is assigned to a NUMA
node. A row interval always has $2^i$ rows for efficiently locating a row
with bit operations. The row interval size is also multiple of the tile size of
a sparse matrix so that multiplication on a tile only needs to access rows
from a single row interval. We store elements in a row interval in row-major
order to increase data locality in SpMM.

\subsubsection{Execution of sparse matrix multiplication} \label{sec:exec}
We perform sparse matrix dense matrix multiplication in semi-external memory
and optimize it for different numbers of columns in the dense matrices.
Thanks to the semi-external memory execution, sparse matrix multiplication
streams data in the sparse matrix from SSDs, which maximizes I/O throughput
of the SSDs.

We partition a sparse matrix horizontally for parallelization and assign multiple
contiguous tile rows to the same partition (Figure \ref{sparse_mat}).
The number of tile rows assigned to a partition is determined at runtime
based on the number of columns in the input dense matrix. A thread reads
a partition of the sparse matrix asynchronously from SSDs. Once a partition
is ready in memory, the worker thread multiplies the partition with the input
dense matrix. A thread processes one tile at a time and stores
the intermediate result in a buffer allocated in the local memory to reduce
remote memory access. 

To better utilize CPU cache, we process tiles of a partition in
\textit{super tile}s (Figure \ref{sparse_mat}). The tile size of a sparse
matrix is specified when the sparse matrix image is created and is relatively
small to handle different numbers of columns in the dense matrices.
A \textit{super tile} is composed of tiles from multiple tile rows and its
size is determined at runtime by three factors: the number of columns
in the dense matrices, the CPU cache size and the number of threads that
share the CPU cache. An optimal size for a \textit{super tile} fills
the CPU cache with the rows from the dense matrices involved in
the computation with the \textit{super tile}.

Load balancing also plays a key role in sparse matrix multiplication on
many real-world graphs due to their power-law distribution in vertex degree.
In FlashEigen, a worker thread first processes partitions originally assigned
to the thread. When a worker thread finishes
all of its own partitions, it steals partitions that have not been processed
from other worker threads.

In spite of nearly random edge connection in a real-world graph,
there exists regularity that allows vectorization to improve performance
in sparse matrix dense matrix multiplication. For each non-zero entry, we
need to multiply it with the corresponding row from the input dense matrix
and add the result to the corresponding row in the output dense matrix.
These operations can be accomplished by the vector CPU instructions such as
AVX \cite{avx}. The current implementation relies on GCC's auto-vectorization
to translate the C code to the vector CPU instructions by predefining the matrix
width in the code.

When accessing a sparse matrix on SSDs, we keep a set of memory buffers for
I/O access to reduce the overhead of memory allocation.
For a large spare matrix, each tile row is fairly large, on the order
of tens of megabytes. The operating system usually allocate a memory buffer
for such an I/O size with \textit{mmap()} and populates the buffer with physical
pages when the buffer is used. It is computationally expensive to populate
large memory buffers frequently. When accessing high-throughput I/O devices,
such overhead can cause substantial performance loss. Therefore, we keep a set
of memory buffers allocated previously and reuse them for new I/O requests.
Because tile rows in a sparse matrix usually have differnt sizes, we resize
a previously allocated memory buffer if it is too small for a new I/O request.

\subsection{The vector subspace}
The vector subspace required by an eigensolver is massive when the eigensolver
computes eigenvalues of a billion-node graph or computes many eigenvalues
of a multi-million-node graph. The number of vectors in the subspace
increases with the number of required eigenvalues. Furthermore, a larger
number of vectors in the subspace potentially improves the convergence rate
of an eigensolver. The storage size required by the subspace is often larger than
the sparse matrix for eigendecomposition on many real-world graphs. Therefore,
FlashEigen stores these vectors on SSDs.

\begin{table}
	\begin{center}
		\small
		\begin{tabular}{|c|c|c|c|c|}
			\hline
			name & operation \\
			\hline
			\textit{MvTimesMatAddMv} & $CC \leftarrow \alpha \times AA \times B + \beta \times CC$ \\
			\hline
			\textit{MvTransMv} & $A \leftarrow \alpha \times t(AA) \times BB$ \\
			\hline
			\textit{MvScale1} & $BB \leftarrow \alpha \times AA$ \\
			\hline
			\textit{MvScale2} & $BB \leftarrow AA \times diag(vec)$ \\
			\hline
			\textit{MvAddMv} & $CC \leftarrow \alpha \times AA + \beta \times BB$ \\
			\hline
			\textit{MvDot} & $vec[i] \leftarrow t(AA[,i]) * BB[,i]$ \\
			\hline
			\textit{MvNorm} & $vec \leftarrow norm\_col(AA)$ \\
			\hline
			\textit{CloneView} & $AA[,idxs]$ \\
			\hline
			\textit{SetBlock} & $AA[,idxs] \leftarrow BB$ \\
			\hline
			\textit{MvRandom} & $AA \leftarrow rand\_init$ \\
			\hline
			\textit{ConvLayout} & $AA \leftarrow conv\_layout(BB)$ \\
			\hline
		\end{tabular}
		\normalsize
	\end{center}
	\caption{The dense matrix operations required by the Anasazi eigensolvers.
		$AA$ and $BB$ represents a tall dense matrix, $A$ and $B$ represents
		a small dense matrix, $\alpha$ and $\beta$ represents scalar variables.}
	\label{anasazi_ops}
\end{table}

FlashEigen implements a set of dense matrix operations shown in Table
\ref{anasazi_ops}. The Anasazi framework provides a set of programming
interfaces that expose the vectors in the subspace to users as dense matrices
and allow users to redefine the dense matrices and the operations on them.
The first ten operations are the ones required by the Anasazi
framework. The most computationally expensive operations are the two
matrix multiplication operations: \textit{MvTimesMatAddMv} and \textit{MvTransMv},
mainly used for reorthogonalization to fix floating-point rounding errors.
The eigensolvers use \textit{CloneView} and \textit{SetBlock} to access individual
columns of a dense matrix, so we store the dense matrices in column major by default.
However, the sparse matrix dense matrix multiplication described in Section
\ref{spmm} requires a row-major dense matrix to increase data locality. Thus,
FlashEigen adds another operation \textit{ConvLayout} to convert data layout
in dense matrices, which converts a column-major matrix to a row-major
matrix when it is passed to the SpMM operation.

External-memory dense matrix operations faces multiple challenges.
Unlike sparse matrix multiplication, these dense matrix operations are less
memory intensive. Even though SSDs are fast, their sequential I/O performance
is still an order of magnitude slower than RAM. Furthermore, SSDs wears out
after a certain amount of write. Even enterprise SSDs \cite{ocz} only allows
a small number of DWPD
(diskful writes per day). Therefore, FlashEigen optimizes dense matrix operations
with three goals: \textit{(i)} maximizing I/O throughput of SSDs, \textit{(ii)}
minimizing the amount of data read from SSDs, \textit{(iii)} reducing SSD
wearout.


\subsubsection{The storage format of the vector subspace}
FlashEigen stores multiple vectors in a dense matrix physically because Anasazi
eigensolvers update multiple vectors of the subspace in an iteration thanks to
the block extension. The number of vectors in a matrix is determined by
the \textit{block size} of a block eigensolver.
As such, the subspace is composed of multiple tall-and-skinny (TAS) dense
matrices.

FlashEigen stores each TAS matrix in a separate SAFS file to leverage
the optimizations in SAFS. This guarantees that I/O accesses to the dense
matrices are evenly distributed to all SSDs by SAFS, regardless of the number
of SSDs and the subspace size. It also eases matrix creation and deletion by
simply creating and deleting an SAFS file.

FlashEigen partitions a TAS matrix horizontally to assist in parallelization
and external-memory access. Figure \ref{dense_mat} (b) illustrates the format
of an external-memory TAS matrix. Like
a NUMA dense matrix in Figure \ref{dense_mat} (a), an external-memory matrix
is divided into multiple row intervals and data in a row interval is stored
contiguously to generate large I/O requests, on the order of megabytes.
The size of a row interval is chosen according to the number of columns in the matrix.
Unlike a NUMA dense matrix, elements in a row interval of an external-memory
matrix are stored in column-major order for easily accessing individual columns.

\subsubsection{Parallelize matrix operations} \label{sec:par}
All large dense matrices in FlashEigen are tall and skinny so we partition
them horizontally for parallelization. All matrix operations in Table
\ref{anasazi_ops} allow a worker thread to process partitions of a matrix
independently.

All of the matrix operations in Table \ref{anasazi_ops} that outputs TAS
matrices are embarrassingly parallelizable on the TAS matrices. In these
operations, a row interval in the output matrix only depends on the same row
interval from all of the TAS input matrices. Thus, the computation and data
access to the TAS matrices are completely independent between row intervals.
To parallelize these matrix operations, we assign one row interval at a time to
a worker thread. When a worker thread gets a row interval, it owns the row
interval and is responsible for accessing the data in the row interval from
all of the TAS matrices and performing computation on them. When processing
a row interval, a worker thread does not need to access data from another row
interval on SSDs.

Some of the operations do not output TAS matrices and their outputs depend on
all row intervals of the input matrices. For example, \textit{MvTransMv}
outputs a small matrix. These operations can usually be split into two
sub-operations: the first one performs computation on each row interval
independently and outputs a small vector or a small matrix; the second one
aggregates all of the small vectors or matrices and outputs a single small
vector or matrix. The first sub-operation accounts for most of computation in
such an operation and requires external-memory data access, so we parallelize
it in the same fashion as the operations that output TAS matrices.


\subsubsection{External memory access} \label{sec:em}
I/O access for dense matrix operations is relatively simple. All matrix partitions
have the same size and each operation requires to access all data in a TAS
matrix, which results in sequential I/O access. For most of the matrix operations
in Table \ref{anasazi_ops}, a worker thread reads data in a row interval
asynchronously and perform computation when all data in the row interval is
ready in memory. In this section, we describe some optimizations that aim at
reducing memory consumption and achieving maximal I/O throughput from SSDs.

Keeping data in a row interval from all of the input TAS matrices potentially
consumes a significant amount of memory if a matrix operation involves in many
TAS matrices.
Operations such as \textit{MvTimesMatAddMv} and \textit{MvTransMv} frequently
take as input many vectors in the subspace, which is stored in multiple TAS
dense matrices physically. The actual number of TAS matrices involved in
the two operations varies between iterations and can grow to as large as several
hundred when an eigensolver computes hundreds of eigenvalues. The storage size
of a row interval is usually configured on the order of tens of megabytes to
achieve I/O throughput from SSDs. When there are hundreds of dense matrices in
the subspace, we may keep a substantial amount of data in memory.
Reading only part of a partition results in many small reads and writes to
SSDs because the matrices are organized in column-major order.

\begin{figure}
\centering
\includegraphics[scale=0.4]{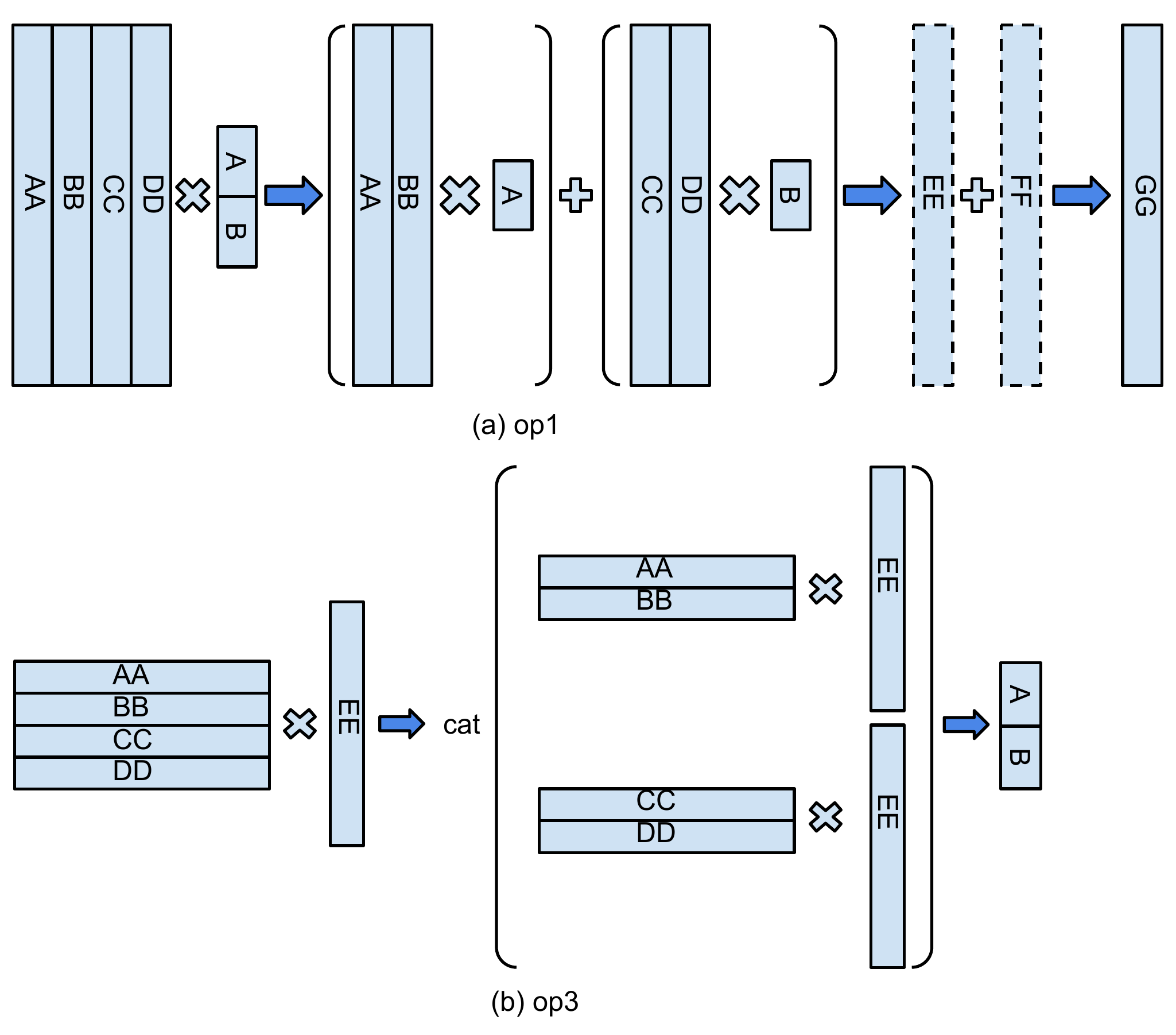}
\vspace{-5pt}
\caption{Break a large group of dense matrices in an operation into multiple
small groups to reduce memory consumption. XX indicates a TAS matrix
stored on SSDs and X indicates a small matrix stored in memory.}
\vspace{-5pt}
\label{fig:mat_group}
\end{figure}

Instead, we break a large group of TAS matrices into multiple small groups to
reduce memory consumption for these matrix operations. Figure \ref{fig:mat_group}
illustrates this optimization on \textit{MvTimesMatAddMv} and \textit{MvTransMv}.
For \textit{MvTimesMatAddMv}, we split the small dense matrix horizontally and
each group of TAS matrices gets a partition of the small dense matrix. Each group
generates an intermediate TAS matrix conceptually and we apply an addition
operation on all of the intermediate matrices to generate the final result.
As such, we perform computation on each group separately and only need to keep
the data in a row interval from all of the TAS matrices in a group. Therefore,
memory consumption is determined by the group size instead of the number of TAS
matrices involved in the operation. Materializing these intermediate matrices
would result in large memory consumption if we store them in memory, or a large
amount of I/O if we store them on SSDs. Instead, we only materialize part of
the intermediate matrices and passes the materialized parts to
the addition operation to generate the final result. We apply a similar strategy
to \textit{MvTransMv}, which requires each group to share the same TAS matrix in
the right operand. Each group generates a small matrix that is small enough to
be kept in memory. To minimize I/O, we share I/O for accessing the TAS matrix
in the right operand.

Like sparse matrix multiplication, we maintain a per-thread memory buffer pool
for I/O access to reduce the overhead of memory allocation when accessing dense
matrices on SSDs. To increase I/O throughput, we access data in dense matrices
on SSDs with large I/O requests. Allocating a large memory buffer for each I/O
request causes the operating system to populate memory with physical pages
when the buffer is used and this is a computationally expensive operation.
Therefore, we maintain a pool of memory buffers
with physical pages populated in advance and associate each I/O request with
a buffer from the set. We maintain such a pool for each worker thread to
minimize locking overhead.

\subsubsection{Matrix caching}
FlashEigen deploys two forms of matrix caching to reduce I/O to SSDs.
In the first case, FlashEigen worker threads cache part of a TAS
matrix locally. In the other case, FlashEigen can also cache the most recent
TAS matrix in the vector subspace.

Caching part of a TAS matrix read from SSDs benefits some of
the matrix operations. One example is the optimization for $op3$ in section
\ref{sec:em}, which breaks a large group of dense matrices into multiple
subgroups and the matrix in the right operand is shared by all of the subgroups.
A worker thread only needs to cache data in a row interval of the right matrix
that is currently being processed because a thread processes one row interval
at a time and each row interval of the TAS matrices is processed only once.
Therefore, a worker thread can buffers the data in a row interval of a matrix
locally and accessing the buffered data doesn't incur any locking overhead.

When we buffer the recent portions, we need to give each matrix a data identifier
to identify its data, so we can recognize which portion of data can be reused.
For certain operations, even though a new matrix is created, the data inside
remains the same. A typical example is transpose. The identifier we give to
each matrix should identify the data inside a matrix instead of individual matrices,
so a transposed matrix and its original matrix should share the same identifier.

FlashEigen also caches the most recent TAS matrix in the vector subspace if
the RAM of a machine is sufficient to accommodate the entire matrix.
When a new matrix in the vector subspace is generated by sparse matrix
multiplication, an eigensolver needs to perform a sequence of operations on it,
which includes reorgonalization. By caching the matrix in memory, we can
significantly reduce the amount of data written to SSDs.

\section{Experimental Evaluation}

We evaluate the performance of the SSD-based FlashEigen on multiple real-world
billion-scale graphs including a web-page graph with 3.4 billion vertices.
We first evaluate the performance
of the two most computationally intensive computation in the eigensolvers:
sparse matrix dense matrix multiplication and dense matrix multiplication.
We demonstrate the effectiveness of the optimizations on the two operations
and compare the performance of our external-memory implementation with
multiple in-memory implementations: \textit{(i)} our in-memory implementations,
\textit{(ii)} MKL and \textit{(iii)} Trilinos. We then evaluate the overall
performance of FlashEigen and compare with the original Anasazi eigensolvers.
We further demonstrate the scalability
of FlashEigen on a web graph of 3.4 billion vertices and 129 billion edges.

We conduct all experiments on a non-uniform memory architecture machine with
four Intel Xeon E7-4860 processors, clocked at 2.6 GHz, and 1TB memory of
DDR3-1600. Each processor has 12 cores. The machine has three LSI SAS 9300-8e
host bus adapters (HBA) connected to a SuperMicro storage chassis, in which
24 OCZ Intrepid 3000 SSDs are installed. The 24 SSDs together are capable of
delivering 12 GB/s for read and 10 GB/s for write at maximum. The machine runs
Linux kernel v3.13.0. We use 48 threads for most of experiments by default.

\begin{table}
\begin{center}
\footnotesize
\begin{tabular}{|c|c|c|c|c|}
\hline
Graph datasets & \# Vertices & \# Edges & Directed \\
\hline
Twitter \cite{twitter} & $42$M & $1.5$B & Yes \\
\hline
Friendster \cite{friendster} & $65$M & $1.7$B & No \\
\hline
KNN distance graph \cite{} & $62$M & $12$B & No \\
\hline
Page \cite{web_graph} & $3.4$B & $129$B & Yes \\
\hline
\end{tabular}
\normalsize
\end{center}
\vspace{-10pt}
\caption{Graph data sets.}
\label{graphs}
\end{table}

We use the real-world graphs in Table \ref{graphs} for evaluation. The largest
graph is the page graph with 3.4 billion vertices and 129 billion edges.
The smallest graph we use has 42 million vertices and 1.5 billion edges.
Twitter and Friendster are social network graphs. The KNN distance graph is
a symmetrized 100-nearest neighbor adjacency graph with cosine distance as
the edge weight. The distance graph is generated over all frames of the Babel
Tagalog corpus, commonly used in speech recognition. Majority of the vertices
in this graph has degree between $100$ to $1000$, so this graph does not follow
the power-law distribution in vertex degree.
The page graph is clustered by domain, generating good CPU cache hit rates
in sparse matrix dense matrix multiplication.

For performance evaluation, we implement an in-memory implementation of both
sparse matrix multiplication and dense matrix multiplication. Therefore, our
FlashEigen eigensolver is able to run both in memory and on SSDs.
In the following sections, we denote the eigensolver in memory by FE-IM and
the eigensolver on SSDs with FM-SEM.

\subsection{Sparse matrix multiplication}
This section evaluates the performance of the semi-external-memory (SEM)
implementation of SpMM in FlashEigen. We first evaluate the effectiveness
of the optimizations on the SpMM. Then we compare its performance with that
of the Intel MKL and Trilinos implementations.

\subsubsection{Optimizations on SpMM}
We first deploy a set of memory optimizations to implement a fast
in-memory sparse matrix multiplication and then deploy a set of I/O
optimizations to further speed up this operation in semi-external memory.
We apply the memory and I/O optimizations incrementally to illustrate their
effectiveness.

For memory optimizations, we start with an implementation that performs sparse
matrix multiplication on a sparse matrix in the CSR format and apply
the optimizations incrementally in the following order:
\begin{itemize} \itemsep1pt \parskip0pt \parsep0pt
		\item partition dense matrices for NUMA (\textit{NUMA}),
	\item partition the sparse matrix in both dimensions into tiles of
		$16K \times 16K$ (\textit{Cache blocking}),
	\item organize multiple physical tiles into a super tile to fill CPU cache
		(\textit{Super tile}),
	\item use CPU vectorization instructions (\textit{Vec}),
	\item allocate a local buffer to store the intermediate result of
		multiplication of tiles and the input dense matrix(\textit{Local write}),
	\item combine the SCSR and COO format to reduce the number of conditional
		jump CPU instructions (\textit{SCSR+COO}),
\end{itemize}

\begin{figure}
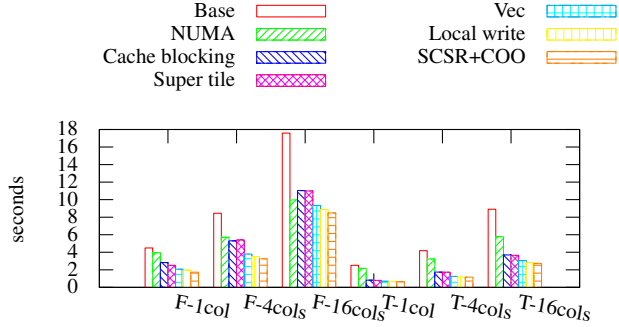

	\begin{center}
		\footnotesize
		\vspace{-15pt}
		\include{SpMM_optimize}
		\vspace{-15pt}
		\caption{The effectiveness of the SpMM optimizations on the Friendster
			graph (F) and the Twitter graph (T) for different numbers of
			columns in the dense matrices.}
		\label{perf:spmm_opt}
	\end{center}
\end{figure}

Figure \ref{perf:spmm_opt} shows that almost all of these optimizations have
positive effect on sparse matrix multiplication and all optimizations
together speed up the operation by $2-4$ times.
The degree of effectiveness varies
significantly between different graphs and different numbers of columns in
the dense matrices. For example, the NUMA optimization is more effective when
the dense matrices have more columns because more columns in the dense matrices
require more memory bandwidth. Cache blocking is very effective when
the dense matrices have fewer columns because it can effectively increase CPU
cache hits. When there are more columns in the dense matrices, data locality
improves and cache blocking becomes less effective. When there are too many
columns, the rows from
the input and output matrices can no longer be in the CPU cache. Thus, it even
has a little negative effect on the Friendster graph when the dense matrices
have $16$ columns. However, we never use dense matrices with more than four
columns in sparse matrix multiplication in the KrylovSchur eigensolver.
With all optimizations, we have a fast in-memory sparse matrix dense matrix
multiplication, denoted by FE-IM SpMM.

\subsubsection{SpMM performance}

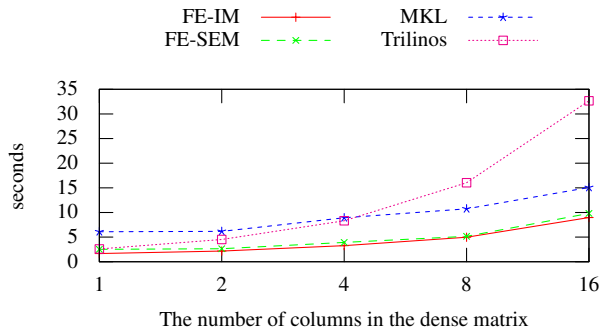
\begin{figure}
	\begin{center}
		\footnotesize
		\vspace{-15pt}
		\begin{tikzpicture}[gnuplot]
\path (0.000,0.000) rectangle (8.382,4.572);
\gpcolor{color=gp lt color border}
\gpsetlinetype{gp lt border}
\gpsetlinewidth{1.00}
\draw[gp path] (1.320,0.985)--(1.500,0.985);
\draw[gp path] (7.829,0.985)--(7.649,0.985);
\node[gp node right] at (1.136,0.985) { 0};
\draw[gp path] (1.320,1.313)--(1.500,1.313);
\draw[gp path] (7.829,1.313)--(7.649,1.313);
\node[gp node right] at (1.136,1.313) { 5};
\draw[gp path] (1.320,1.640)--(1.500,1.640);
\draw[gp path] (7.829,1.640)--(7.649,1.640);
\node[gp node right] at (1.136,1.640) { 10};
\draw[gp path] (1.320,1.968)--(1.500,1.968);
\draw[gp path] (7.829,1.968)--(7.649,1.968);
\node[gp node right] at (1.136,1.968) { 15};
\draw[gp path] (1.320,2.296)--(1.500,2.296);
\draw[gp path] (7.829,2.296)--(7.649,2.296);
\node[gp node right] at (1.136,2.296) { 20};
\draw[gp path] (1.320,2.624)--(1.500,2.624);
\draw[gp path] (7.829,2.624)--(7.649,2.624);
\node[gp node right] at (1.136,2.624) { 25};
\draw[gp path] (1.320,2.951)--(1.500,2.951);
\draw[gp path] (7.829,2.951)--(7.649,2.951);
\node[gp node right] at (1.136,2.951) { 30};
\draw[gp path] (1.320,3.279)--(1.500,3.279);
\draw[gp path] (7.829,3.279)--(7.649,3.279);
\node[gp node right] at (1.136,3.279) { 35};
\draw[gp path] (1.320,0.985)--(1.320,1.165);
\draw[gp path] (1.320,3.279)--(1.320,3.099);
\node[gp node center] at (1.320,0.677) {1};
\draw[gp path] (2.947,0.985)--(2.947,1.165);
\draw[gp path] (2.947,3.279)--(2.947,3.099);
\node[gp node center] at (2.947,0.677) {2};
\draw[gp path] (4.575,0.985)--(4.575,1.165);
\draw[gp path] (4.575,3.279)--(4.575,3.099);
\node[gp node center] at (4.575,0.677) {4};
\draw[gp path] (6.202,0.985)--(6.202,1.165);
\draw[gp path] (6.202,3.279)--(6.202,3.099);
\node[gp node center] at (6.202,0.677) {8};
\draw[gp path] (7.829,0.985)--(7.829,1.165);
\draw[gp path] (7.829,3.279)--(7.829,3.099);
\node[gp node center] at (7.829,0.677) {16};
\draw[gp path] (1.320,3.279)--(1.320,0.985)--(7.829,0.985)--(7.829,3.279)--cycle;
\node[gp node center,rotate=-270] at (0.246,2.132) {seconds};
\node[gp node center] at (4.574,0.215) {The number of columns in the dense matrix};
\node[gp node right] at (3.290,4.238) {FE-IM};
\gpcolor{color=gp lt color 0}
\gpsetlinetype{gp lt plot 0}
\draw[gp path] (3.474,4.238)--(4.390,4.238);
\draw[gp path] (1.320,1.094)--(2.947,1.127)--(4.575,1.199)--(6.202,1.311)--(7.829,1.573);
\gpsetpointsize{4.00}
\gppoint{gp mark 1}{(1.320,1.094)}
\gppoint{gp mark 1}{(2.947,1.127)}
\gppoint{gp mark 1}{(4.575,1.199)}
\gppoint{gp mark 1}{(6.202,1.311)}
\gppoint{gp mark 1}{(7.829,1.573)}
\gppoint{gp mark 1}{(3.932,4.238)}
\gpcolor{color=gp lt color border}
\node[gp node right] at (3.290,3.930) {FE-SEM};
\gpcolor{color=gp lt color 1}
\gpsetlinetype{gp lt plot 1}
\draw[gp path] (3.474,3.930)--(4.390,3.930);
\draw[gp path] (1.320,1.150)--(2.947,1.158)--(4.575,1.242)--(6.202,1.323)--(7.829,1.628);
\gppoint{gp mark 2}{(1.320,1.150)}
\gppoint{gp mark 2}{(2.947,1.158)}
\gppoint{gp mark 2}{(4.575,1.242)}
\gppoint{gp mark 2}{(6.202,1.323)}
\gppoint{gp mark 2}{(7.829,1.628)}
\gppoint{gp mark 2}{(3.932,3.930)}
\gpcolor{color=gp lt color border}
\node[gp node right] at (6.046,4.238) {MKL};
\gpcolor{color=gp lt color 2}
\gpsetlinetype{gp lt plot 2}
\draw[gp path] (6.230,4.238)--(7.146,4.238);
\draw[gp path] (1.320,1.384)--(2.947,1.389)--(4.575,1.570)--(6.202,1.689)--(7.829,1.973);
\gppoint{gp mark 3}{(1.320,1.384)}
\gppoint{gp mark 3}{(2.947,1.389)}
\gppoint{gp mark 3}{(4.575,1.570)}
\gppoint{gp mark 3}{(6.202,1.689)}
\gppoint{gp mark 3}{(7.829,1.973)}
\gppoint{gp mark 3}{(6.688,4.238)}
\gpcolor{color=gp lt color border}
\node[gp node right] at (6.046,3.930) {Trilinos};
\gpcolor{color=gp lt color 3}
\gpsetlinetype{gp lt plot 3}
\draw[gp path] (6.230,3.930)--(7.146,3.930);
\draw[gp path] (1.320,1.155)--(2.947,1.281)--(4.575,1.530)--(6.202,2.036)--(7.829,3.126);
\gppoint{gp mark 4}{(1.320,1.155)}
\gppoint{gp mark 4}{(2.947,1.281)}
\gppoint{gp mark 4}{(4.575,1.530)}
\gppoint{gp mark 4}{(6.202,2.036)}
\gppoint{gp mark 4}{(7.829,3.126)}
\gppoint{gp mark 4}{(6.688,3.930)}
\gpcolor{color=gp lt color border}
\gpsetlinetype{gp lt border}
\draw[gp path] (1.320,3.279)--(1.320,0.985)--(7.829,0.985)--(7.829,3.279)--cycle;
\gpdefrectangularnode{gp plot 1}{\pgfpoint{1.320cm}{0.985cm}}{\pgfpoint{7.829cm}{3.279cm}}
\end{tikzpicture}
		\vspace{-15pt}
		\caption{The runtime of in-memory SpMM (FE-IM) and SEM-SpMM (FE-SEM)
			in the FlashEigen, the MKL and the Trilinos implementation on
		the Friendster graph.}
		\label{perf:spmm}
	\end{center}
\end{figure}

We then evaluate the performance of the SEM SpMM in FlashEigen and compare its
performance with FE-IM SpMM, the MKL implementation and the Trilinos
implementation (Figure \ref{perf:spmm}). We only show the performance of
SpMM on the Friendster graph and the performance on the other graphs is
similar. The MKL and Trilinos SpMM cannot run on the page graph.

Our SEM SpMM has performance comparable to FE-IM SpMM, while both FE-IM SpMM
and FE-SEM SpMM outperforms the MKL and Trilinos implementations.
On the Friendster graph, FE-SEM SpMM achieves 60\% performance of FE-IM SpMM
when the dense matrix has only one column and the performance gap narrows
as the number of columns in the dense matrices increases.
The Trilinos SpMM is optimized for sparse matrix vector multiplication (SpMV).
But even for SpMV, our IM-SpMM outperforms Trilinos by 36\%. The MKL SpMM
performs better when the dense matrices have more columns, but our
implementations can still outperform MKL by $2-3$ times in most settings.

\begin{figure}[t]
\centering
\footnotesize
\vspace{-15pt}
\begin{subfigure}{.5\linewidth}
	\begin{tikzpicture}[gnuplot]
\path (0.000,0.000) rectangle (5.588,4.572);
\gpcolor{color=gp lt color border}
\gpsetlinetype{gp lt border}
\gpsetlinewidth{1.00}
\draw[gp path] (1.196,1.126)--(1.376,1.126);
\draw[gp path] (4.345,1.126)--(4.165,1.126);
\node[gp node right] at (1.012,1.126) { 0};
\draw[gp path] (1.196,1.557)--(1.376,1.557);
\draw[gp path] (4.345,1.557)--(4.165,1.557);
\node[gp node right] at (1.012,1.557) { 0.2};
\draw[gp path] (1.196,1.987)--(1.376,1.987);
\draw[gp path] (4.345,1.987)--(4.165,1.987);
\node[gp node right] at (1.012,1.987) { 0.4};
\draw[gp path] (1.196,2.418)--(1.376,2.418);
\draw[gp path] (4.345,2.418)--(4.165,2.418);
\node[gp node right] at (1.012,2.418) { 0.6};
\draw[gp path] (1.196,2.848)--(1.376,2.848);
\draw[gp path] (4.345,2.848)--(4.165,2.848);
\node[gp node right] at (1.012,2.848) { 0.8};
\draw[gp path] (1.196,3.279)--(1.376,3.279);
\draw[gp path] (4.345,3.279)--(4.165,3.279);
\node[gp node right] at (1.012,3.279) { 1};
\draw[gp path] (1.721,1.126)--(1.721,1.306);
\draw[gp path] (1.721,3.279)--(1.721,3.099);
\node[gp node left,rotate=-20] at (1.721,0.942) {rmat-100M-40};
\draw[gp path] (2.246,1.126)--(2.246,1.306);
\draw[gp path] (2.246,3.279)--(2.246,3.099);
\node[gp node left,rotate=-20] at (2.246,0.942) {rmat-100M-160};
\draw[gp path] (2.771,1.126)--(2.771,1.306);
\draw[gp path] (2.771,3.279)--(2.771,3.099);
\node[gp node left,rotate=-20] at (2.771,0.942) {W};
\draw[gp path] (3.295,1.126)--(3.295,1.306);
\draw[gp path] (3.295,3.279)--(3.295,3.099);
\node[gp node left,rotate=-20] at (3.295,0.942) {Friendster};
\draw[gp path] (3.820,1.126)--(3.820,1.306);
\draw[gp path] (3.820,3.279)--(3.820,3.099);
\node[gp node left,rotate=-20] at (3.820,0.942) {Page graph};
\draw[gp path] (1.196,3.279)--(1.196,1.126)--(4.345,1.126)--(4.345,3.279)--cycle;
\node[gp node right] at (2.864,4.238) {Trilinos};
\def\gpfillpath{(3.048,4.161)--(3.964,4.161)--(3.964,4.315)--(3.048,4.315)--cycle}
\gpfill{color=gpbgfillcolor} \gpfillpath;
\gpfill{color=gp lt color 0,gp pattern 0,pattern color=.} \gpfillpath;
\gpcolor{color=gp lt color 0}
\gpsetlinetype{gp lt plot 0}
\draw[gp path] (3.048,4.161)--(3.964,4.161)--(3.964,4.315)--(3.048,4.315)--cycle;
\def\gpfillpath{(1.655,1.126)--(1.787,1.126)--(1.787,2.526)--(1.655,2.526)--cycle}
\gpfill{color=gpbgfillcolor} \gpfillpath;
\gpfill{color=gp lt color 0,gp pattern 0,pattern color=.} \gpfillpath;
\draw[gp path] (1.655,1.126)--(1.655,2.525)--(1.786,2.525)--(1.786,1.126)--cycle;
\def\gpfillpath{(2.180,1.126)--(2.312,1.126)--(2.312,2.182)--(2.180,2.182)--cycle}
\gpfill{color=gpbgfillcolor} \gpfillpath;
\gpfill{color=gp lt color 0,gp pattern 0,pattern color=.} \gpfillpath;
\draw[gp path] (2.180,1.126)--(2.180,2.181)--(2.311,2.181)--(2.311,1.126)--cycle;
\def\gpfillpath{(2.705,1.126)--(2.837,1.126)--(2.837,2.419)--(2.705,2.419)--cycle}
\gpfill{color=gpbgfillcolor} \gpfillpath;
\gpfill{color=gp lt color 0,gp pattern 0,pattern color=.} \gpfillpath;
\draw[gp path] (2.705,1.126)--(2.705,2.418)--(2.836,2.418)--(2.836,1.126)--cycle;
\def\gpfillpath{(3.230,1.126)--(3.362,1.126)--(3.362,2.397)--(3.230,2.397)--cycle}
\gpfill{color=gpbgfillcolor} \gpfillpath;
\gpfill{color=gp lt color 0,gp pattern 0,pattern color=.} \gpfillpath;
\draw[gp path] (3.230,1.126)--(3.230,2.396)--(3.361,2.396)--(3.361,1.126)--cycle;
\gpcolor{color=gp lt color border}
\node[gp node right] at (2.864,3.930) {SEM};
\def\gpfillpath{(3.048,3.853)--(3.964,3.853)--(3.964,4.007)--(3.048,4.007)--cycle}
\gpfill{color=gpbgfillcolor} \gpfillpath;
\gpfill{color=gp lt color 1,gp pattern 1,pattern color=.} \gpfillpath;
\gpcolor{color=gp lt color 1}
\gpsetlinetype{gp lt plot 1}
\draw[gp path] (3.048,3.853)--(3.964,3.853)--(3.964,4.007)--(3.048,4.007)--cycle;
\def\gpfillpath{(1.786,1.126)--(1.919,1.126)--(1.919,2.957)--(1.786,2.957)--cycle}
\gpfill{color=gpbgfillcolor} \gpfillpath;
\gpfill{color=gp lt color 1,gp pattern 1,pattern color=.} \gpfillpath;
\draw[gp path] (1.786,1.126)--(1.786,2.956)--(1.918,2.956)--(1.918,1.126)--cycle;
\def\gpfillpath{(2.311,1.126)--(2.443,1.126)--(2.443,2.634)--(2.311,2.634)--cycle}
\gpfill{color=gpbgfillcolor} \gpfillpath;
\gpfill{color=gp lt color 1,gp pattern 1,pattern color=.} \gpfillpath;
\draw[gp path] (2.311,1.126)--(2.311,2.633)--(2.442,2.633)--(2.442,1.126)--cycle;
\def\gpfillpath{(2.836,1.126)--(2.968,1.126)--(2.968,2.204)--(2.836,2.204)--cycle}
\gpfill{color=gpbgfillcolor} \gpfillpath;
\gpfill{color=gp lt color 1,gp pattern 1,pattern color=.} \gpfillpath;
\draw[gp path] (2.836,1.126)--(2.836,2.203)--(2.967,2.203)--(2.967,1.126)--cycle;
\def\gpfillpath{(3.361,1.126)--(3.493,1.126)--(3.493,2.419)--(3.361,2.419)--cycle}
\gpfill{color=gpbgfillcolor} \gpfillpath;
\gpfill{color=gp lt color 1,gp pattern 1,pattern color=.} \gpfillpath;
\draw[gp path] (3.361,1.126)--(3.361,2.418)--(3.492,2.418)--(3.492,1.126)--cycle;
\def\gpfillpath{(3.886,1.126)--(4.018,1.126)--(4.018,2.376)--(3.886,2.376)--cycle}
\gpfill{color=gpbgfillcolor} \gpfillpath;
\gpfill{color=gp lt color 1,gp pattern 1,pattern color=.} \gpfillpath;
\draw[gp path] (3.886,1.126)--(3.886,2.375)--(4.017,2.375)--(4.017,1.126)--cycle;
\gpcolor{color=gp lt color border}
\gpsetlinetype{gp lt border}
\draw[gp path] (1.196,3.279)--(1.196,1.126)--(4.345,1.126)--(4.345,3.279)--cycle;
\gpdefrectangularnode{gp plot 1}{\pgfpoint{1.196cm}{1.126cm}}{\pgfpoint{4.345cm}{3.279cm}}
\end{tikzpicture}
	\vspace{-15pt}
	\caption{SpMV}
	\label{fig:spmm1}
\end{subfigure}%
\begin{subfigure}{.5\linewidth}
	\begin{tikzpicture}[gnuplot]
\path (0.000,0.000) rectangle (5.588,4.572);
\gpcolor{color=gp lt color border}
\gpsetlinetype{gp lt border}
\gpsetlinewidth{1.00}
\draw[gp path] (1.196,1.126)--(1.376,1.126);
\draw[gp path] (4.345,1.126)--(4.165,1.126);
\node[gp node right] at (1.012,1.126) { 0};
\draw[gp path] (1.196,1.557)--(1.376,1.557);
\draw[gp path] (4.345,1.557)--(4.165,1.557);
\node[gp node right] at (1.012,1.557) { 0.2};
\draw[gp path] (1.196,1.987)--(1.376,1.987);
\draw[gp path] (4.345,1.987)--(4.165,1.987);
\node[gp node right] at (1.012,1.987) { 0.4};
\draw[gp path] (1.196,2.418)--(1.376,2.418);
\draw[gp path] (4.345,2.418)--(4.165,2.418);
\node[gp node right] at (1.012,2.418) { 0.6};
\draw[gp path] (1.196,2.848)--(1.376,2.848);
\draw[gp path] (4.345,2.848)--(4.165,2.848);
\node[gp node right] at (1.012,2.848) { 0.8};
\draw[gp path] (1.196,3.279)--(1.376,3.279);
\draw[gp path] (4.345,3.279)--(4.165,3.279);
\node[gp node right] at (1.012,3.279) { 1};
\draw[gp path] (1.721,1.126)--(1.721,1.306);
\draw[gp path] (1.721,3.279)--(1.721,3.099);
\node[gp node left,rotate=-20] at (1.721,0.942) {rmat-100M-40};
\draw[gp path] (2.246,1.126)--(2.246,1.306);
\draw[gp path] (2.246,3.279)--(2.246,3.099);
\node[gp node left,rotate=-20] at (2.246,0.942) {rmat-100M-160};
\draw[gp path] (2.771,1.126)--(2.771,1.306);
\draw[gp path] (2.771,3.279)--(2.771,3.099);
\node[gp node left,rotate=-20] at (2.771,0.942) {W};
\draw[gp path] (3.295,1.126)--(3.295,1.306);
\draw[gp path] (3.295,3.279)--(3.295,3.099);
\node[gp node left,rotate=-20] at (3.295,0.942) {Friendster};
\draw[gp path] (3.820,1.126)--(3.820,1.306);
\draw[gp path] (3.820,3.279)--(3.820,3.099);
\node[gp node left,rotate=-20] at (3.820,0.942) {Page graph};
\draw[gp path] (1.196,3.279)--(1.196,1.126)--(4.345,1.126)--(4.345,3.279)--cycle;
\node[gp node right] at (2.864,4.238) {Trilinos};
\def\gpfillpath{(3.048,4.161)--(3.964,4.161)--(3.964,4.315)--(3.048,4.315)--cycle}
\gpfill{color=gpbgfillcolor} \gpfillpath;
\gpfill{color=gp lt color 0,gp pattern 0,pattern color=.} \gpfillpath;
\gpcolor{color=gp lt color 0}
\gpsetlinetype{gp lt plot 0}
\draw[gp path] (3.048,4.161)--(3.964,4.161)--(3.964,4.315)--(3.048,4.315)--cycle;
\def\gpfillpath{(1.655,1.126)--(1.787,1.126)--(1.787,2.010)--(1.655,2.010)--cycle}
\gpfill{color=gpbgfillcolor} \gpfillpath;
\gpfill{color=gp lt color 0,gp pattern 0,pattern color=.} \gpfillpath;
\draw[gp path] (1.655,1.126)--(1.655,2.009)--(1.786,2.009)--(1.786,1.126)--cycle;
\def\gpfillpath{(2.180,1.126)--(2.312,1.126)--(2.312,1.881)--(2.180,1.881)--cycle}
\gpfill{color=gpbgfillcolor} \gpfillpath;
\gpfill{color=gp lt color 0,gp pattern 0,pattern color=.} \gpfillpath;
\draw[gp path] (2.180,1.126)--(2.180,1.880)--(2.311,1.880)--(2.311,1.126)--cycle;
\def\gpfillpath{(2.705,1.126)--(2.837,1.126)--(2.837,2.053)--(2.705,2.053)--cycle}
\gpfill{color=gpbgfillcolor} \gpfillpath;
\gpfill{color=gp lt color 0,gp pattern 0,pattern color=.} \gpfillpath;
\draw[gp path] (2.705,1.126)--(2.705,2.052)--(2.836,2.052)--(2.836,1.126)--cycle;
\def\gpfillpath{(3.230,1.126)--(3.362,1.126)--(3.362,2.031)--(3.230,2.031)--cycle}
\gpfill{color=gpbgfillcolor} \gpfillpath;
\gpfill{color=gp lt color 0,gp pattern 0,pattern color=.} \gpfillpath;
\draw[gp path] (3.230,1.126)--(3.230,2.030)--(3.361,2.030)--(3.361,1.126)--cycle;
\gpcolor{color=gp lt color border}
\node[gp node right] at (2.864,3.930) {SEM};
\def\gpfillpath{(3.048,3.853)--(3.964,3.853)--(3.964,4.007)--(3.048,4.007)--cycle}
\gpfill{color=gpbgfillcolor} \gpfillpath;
\gpfill{color=gp lt color 1,gp pattern 1,pattern color=.} \gpfillpath;
\gpcolor{color=gp lt color 1}
\gpsetlinetype{gp lt plot 1}
\draw[gp path] (3.048,3.853)--(3.964,3.853)--(3.964,4.007)--(3.048,4.007)--cycle;
\def\gpfillpath{(1.786,1.126)--(1.919,1.126)--(1.919,2.979)--(1.786,2.979)--cycle}
\gpfill{color=gpbgfillcolor} \gpfillpath;
\gpfill{color=gp lt color 1,gp pattern 1,pattern color=.} \gpfillpath;
\draw[gp path] (1.786,1.126)--(1.786,2.978)--(1.918,2.978)--(1.918,1.126)--cycle;
\def\gpfillpath{(2.311,1.126)--(2.443,1.126)--(2.443,2.828)--(2.311,2.828)--cycle}
\gpfill{color=gpbgfillcolor} \gpfillpath;
\gpfill{color=gp lt color 1,gp pattern 1,pattern color=.} \gpfillpath;
\draw[gp path] (2.311,1.126)--(2.311,2.827)--(2.442,2.827)--(2.442,1.126)--cycle;
\def\gpfillpath{(2.836,1.126)--(2.968,1.126)--(2.968,2.785)--(2.836,2.785)--cycle}
\gpfill{color=gpbgfillcolor} \gpfillpath;
\gpfill{color=gp lt color 1,gp pattern 1,pattern color=.} \gpfillpath;
\draw[gp path] (2.836,1.126)--(2.836,2.784)--(2.967,2.784)--(2.967,1.126)--cycle;
\def\gpfillpath{(3.361,1.126)--(3.493,1.126)--(3.493,3.043)--(3.361,3.043)--cycle}
\gpfill{color=gpbgfillcolor} \gpfillpath;
\gpfill{color=gp lt color 1,gp pattern 1,pattern color=.} \gpfillpath;
\draw[gp path] (3.361,1.126)--(3.361,3.042)--(3.492,3.042)--(3.492,1.126)--cycle;
\def\gpfillpath{(3.886,1.126)--(4.018,1.126)--(4.018,2.806)--(3.886,2.806)--cycle}
\gpfill{color=gpbgfillcolor} \gpfillpath;
\gpfill{color=gp lt color 1,gp pattern 1,pattern color=.} \gpfillpath;
\draw[gp path] (3.886,1.126)--(3.886,2.805)--(4.017,2.805)--(4.017,1.126)--cycle;
\gpcolor{color=gp lt color border}
\gpsetlinetype{gp lt border}
\draw[gp path] (1.196,3.279)--(1.196,1.126)--(4.345,1.126)--(4.345,3.279)--cycle;
\gpdefrectangularnode{gp plot 1}{\pgfpoint{1.196cm}{1.126cm}}{\pgfpoint{4.345cm}{3.279cm}}
\end{tikzpicture}
	\vspace{-15pt}
	\caption{SpMM}
	\label{fig:spmm4}
\end{subfigure}
\caption{The performance of Trilinos and FlashEigen-SEM sparse matrix
multiplication relative to FlashEigen-IM sparse matrix multiplication.
In sparse matrix dense multiplication (SpMM), there are four columns
in the dense matrix.}
\vspace{-15pt}
\label{fig:spmm}
\end{figure}
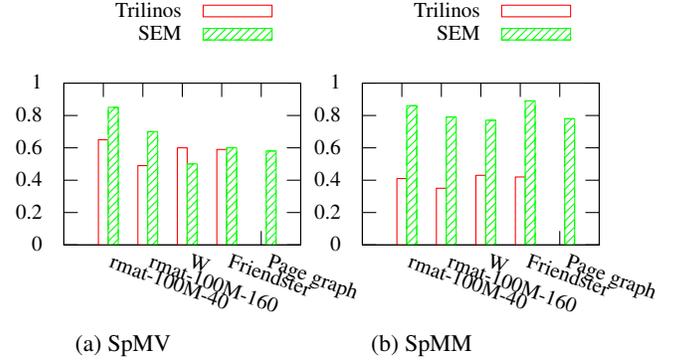


\subsection{Dense matrix multiplication}

This section evaluates the performance of dense matrix multiplication,
the other computationally intensive operation in an eigensolver.
For dense matrix multiplication, we focus on I/O optimizations, so we evaluate
the effectiveness of the I/O optimizations on this operation. Then we
compare the performance of our in-memory and external-memory
implementations with the ones in Intel MKL and Trilinos.

In eigendecomposition, there are two forms of dense matrix multiplication.
The first form, shown by $op1$ in Table \ref{anasazi_ops}, multiplies
a tall-and-skinny dense matrix of $n \times m$
with a small dense matrix of $m \times b$ and outputs a tall-and-skinny dense
matrix of $n \times b$, where $n$ is the size of the eigenvalue problem,
$m$ is the number of existing vectors in the subspace and $b$ is the block
size of the eigensolver. The second form, shown by $op3$, multiplies
a transpose of a tall-and-skinny dense matrix of $n \times m$ with another
tall-and-skinny dense matrix of $n \times b$ and outputs a small matrix.
In both forms, $m$ varies from one block size to the maximal subspace size
specified by a user and is increased by one block size in each
iteration. In the experiments, we set $n$ to 60M, roughly the size of
eigenvalue problems in Table \ref{graphs}, $b$ to 4 and vary $m$
from $4$ to $512$. This setting of $b$ and $m$ is used in the experiments
of our external-memory KrylovSchur eigensolver in the next section.

\subsubsection{Optimizations on dense matrix multiplication}

We illustrate the most effective I/O optimizations on dense matrix
multiplication. We start with a basic implementation of matrix multiplication
in SAFS and deploy the optimizations on SAFS and matrix multiplication
incrementally in the following order:
\begin{itemize} \itemsep1pt \parskip0pt \parsep0pt
	\item use different random striping orders for each file (\textit{diff strip}),
	\item use a per-thread buffer pool to allocate memory for I/O (\textit{buf pool}),
	\item use one I/O thread per NUMA node (\textit{1IOT}),
	\item use I/O polling in worker threads (\textit{polling}),
	\item use 8MB for the maximal block size in the kernel (\textit{max block}),
\end{itemize}

\begin{figure}
	\begin{center}
		\footnotesize
		\vspace{-15pt}
		\begin{tikzpicture}[gnuplot]
\path (0.000,0.000) rectangle (8.382,4.572);
\gpcolor{color=gp lt color border}
\gpsetlinetype{gp lt border}
\gpsetlinewidth{1.00}
\draw[gp path] (1.320,0.985)--(1.500,0.985);
\draw[gp path] (7.829,0.985)--(7.649,0.985);
\node[gp node right] at (1.136,0.985) { 0};
\draw[gp path] (1.320,1.206)--(1.500,1.206);
\draw[gp path] (7.829,1.206)--(7.649,1.206);
\node[gp node right] at (1.136,1.206) { 5};
\draw[gp path] (1.320,1.426)--(1.500,1.426);
\draw[gp path] (7.829,1.426)--(7.649,1.426);
\node[gp node right] at (1.136,1.426) { 10};
\draw[gp path] (1.320,1.647)--(1.500,1.647);
\draw[gp path] (7.829,1.647)--(7.649,1.647);
\node[gp node right] at (1.136,1.647) { 15};
\draw[gp path] (1.320,1.868)--(1.500,1.868);
\draw[gp path] (7.829,1.868)--(7.649,1.868);
\node[gp node right] at (1.136,1.868) { 20};
\draw[gp path] (1.320,2.088)--(1.500,2.088);
\draw[gp path] (7.829,2.088)--(7.649,2.088);
\node[gp node right] at (1.136,2.088) { 25};
\draw[gp path] (1.320,2.309)--(1.500,2.309);
\draw[gp path] (7.829,2.309)--(7.649,2.309);
\node[gp node right] at (1.136,2.309) { 30};
\draw[gp path] (1.320,2.530)--(1.500,2.530);
\draw[gp path] (7.829,2.530)--(7.649,2.530);
\node[gp node right] at (1.136,2.530) { 35};
\draw[gp path] (1.320,2.750)--(1.500,2.750);
\draw[gp path] (7.829,2.750)--(7.649,2.750);
\node[gp node right] at (1.136,2.750) { 40};
\draw[gp path] (1.320,2.971)--(1.500,2.971);
\draw[gp path] (7.829,2.971)--(7.649,2.971);
\node[gp node right] at (1.136,2.971) { 45};
\draw[gp path] (3.490,0.985)--(3.490,1.165);
\draw[gp path] (3.490,2.971)--(3.490,2.791);
\node[gp node center] at (3.490,0.677) {64};
\draw[gp path] (5.659,0.985)--(5.659,1.165);
\draw[gp path] (5.659,2.971)--(5.659,2.791);
\node[gp node center] at (5.659,0.677) {256};
\draw[gp path] (1.320,2.971)--(1.320,0.985)--(7.829,0.985)--(7.829,2.971)--cycle;
\node[gp node center,rotate=-270] at (0.246,1.978) {seconds};
\node[gp node center] at (4.574,0.215) {The number of columns in the left matrix};
\node[gp node right] at (3.290,4.238) {Base};
\def\gpfillpath{(3.474,4.161)--(4.390,4.161)--(4.390,4.315)--(3.474,4.315)--cycle}
\gpfill{color=gpbgfillcolor} \gpfillpath;
\gpfill{color=gp lt color 0,gp pattern 0,pattern color=.} \gpfillpath;
\gpcolor{color=gp lt color 0}
\gpsetlinetype{gp lt plot 0}
\draw[gp path] (3.474,4.161)--(4.390,4.161)--(4.390,4.315)--(3.474,4.315)--cycle;
\def\gpfillpath{(2.812,0.985)--(3.084,0.985)--(3.084,1.483)--(2.812,1.483)--cycle}
\gpfill{color=gpbgfillcolor} \gpfillpath;
\gpfill{color=gp lt color 0,gp pattern 0,pattern color=.} \gpfillpath;
\draw[gp path] (2.812,0.985)--(2.812,1.482)--(3.083,1.482)--(3.083,0.985)--cycle;
\def\gpfillpath{(4.981,0.985)--(5.254,0.985)--(5.254,2.766)--(4.981,2.766)--cycle}
\gpfill{color=gpbgfillcolor} \gpfillpath;
\gpfill{color=gp lt color 0,gp pattern 0,pattern color=.} \gpfillpath;
\draw[gp path] (4.981,0.985)--(4.981,2.765)--(5.253,2.765)--(5.253,0.985)--cycle;
\gpcolor{color=gp lt color border}
\node[gp node right] at (3.290,3.930) {diff stripe};
\def\gpfillpath{(3.474,3.853)--(4.390,3.853)--(4.390,4.007)--(3.474,4.007)--cycle}
\gpfill{color=gpbgfillcolor} \gpfillpath;
\gpfill{color=gp lt color 1,gp pattern 1,pattern color=.} \gpfillpath;
\gpcolor{color=gp lt color 1}
\gpsetlinetype{gp lt plot 1}
\draw[gp path] (3.474,3.853)--(4.390,3.853)--(4.390,4.007)--(3.474,4.007)--cycle;
\def\gpfillpath{(3.083,0.985)--(3.355,0.985)--(3.355,1.446)--(3.083,1.446)--cycle}
\gpfill{color=gpbgfillcolor} \gpfillpath;
\gpfill{color=gp lt color 1,gp pattern 1,pattern color=.} \gpfillpath;
\draw[gp path] (3.083,0.985)--(3.083,1.445)--(3.354,1.445)--(3.354,0.985)--cycle;
\def\gpfillpath{(5.253,0.985)--(5.525,0.985)--(5.525,2.660)--(5.253,2.660)--cycle}
\gpfill{color=gpbgfillcolor} \gpfillpath;
\gpfill{color=gp lt color 1,gp pattern 1,pattern color=.} \gpfillpath;
\draw[gp path] (5.253,0.985)--(5.253,2.659)--(5.524,2.659)--(5.524,0.985)--cycle;
\gpcolor{color=gp lt color border}
\node[gp node right] at (3.290,3.622) {buf pool};
\def\gpfillpath{(3.474,3.545)--(4.390,3.545)--(4.390,3.699)--(3.474,3.699)--cycle}
\gpfill{color=gpbgfillcolor} \gpfillpath;
\gpfill{color=gp lt color 2,gp pattern 2,pattern color=.} \gpfillpath;
\gpcolor{color=gp lt color 2}
\gpsetlinetype{gp lt plot 2}
\draw[gp path] (3.474,3.545)--(4.390,3.545)--(4.390,3.699)--(3.474,3.699)--cycle;
\def\gpfillpath{(3.354,0.985)--(3.626,0.985)--(3.626,1.273)--(3.354,1.273)--cycle}
\gpfill{color=gpbgfillcolor} \gpfillpath;
\gpfill{color=gp lt color 2,gp pattern 2,pattern color=.} \gpfillpath;
\draw[gp path] (3.354,0.985)--(3.354,1.272)--(3.625,1.272)--(3.625,0.985)--cycle;
\def\gpfillpath{(5.524,0.985)--(5.796,0.985)--(5.796,2.253)--(5.524,2.253)--cycle}
\gpfill{color=gpbgfillcolor} \gpfillpath;
\gpfill{color=gp lt color 2,gp pattern 2,pattern color=.} \gpfillpath;
\draw[gp path] (5.524,0.985)--(5.524,2.252)--(5.795,2.252)--(5.795,0.985)--cycle;
\gpcolor{color=gp lt color border}
\node[gp node right] at (6.598,4.238) {1IOT};
\def\gpfillpath{(6.782,4.161)--(7.698,4.161)--(7.698,4.315)--(6.782,4.315)--cycle}
\gpfill{color=gpbgfillcolor} \gpfillpath;
\gpfill{color=gp lt color 3,gp pattern 3,pattern color=.} \gpfillpath;
\gpcolor{color=gp lt color 3}
\gpsetlinetype{gp lt plot 3}
\draw[gp path] (6.782,4.161)--(7.698,4.161)--(7.698,4.315)--(6.782,4.315)--cycle;
\def\gpfillpath{(3.625,0.985)--(3.897,0.985)--(3.897,1.182)--(3.625,1.182)--cycle}
\gpfill{color=gpbgfillcolor} \gpfillpath;
\gpfill{color=gp lt color 3,gp pattern 3,pattern color=.} \gpfillpath;
\draw[gp path] (3.625,0.985)--(3.625,1.181)--(3.896,1.181)--(3.896,0.985)--cycle;
\def\gpfillpath{(5.795,0.985)--(6.067,0.985)--(6.067,1.769)--(5.795,1.769)--cycle}
\gpfill{color=gpbgfillcolor} \gpfillpath;
\gpfill{color=gp lt color 3,gp pattern 3,pattern color=.} \gpfillpath;
\draw[gp path] (5.795,0.985)--(5.795,1.768)--(6.066,1.768)--(6.066,0.985)--cycle;
\gpcolor{color=gp lt color border}
\node[gp node right] at (6.598,3.930) {polling};
\def\gpfillpath{(6.782,3.853)--(7.698,3.853)--(7.698,4.007)--(6.782,4.007)--cycle}
\gpfill{color=gpbgfillcolor} \gpfillpath;
\gpfill{color=gp lt color 4,gp pattern 4,pattern color=.} \gpfillpath;
\gpcolor{color=gp lt color 4}
\gpsetlinetype{gp lt plot 4}
\draw[gp path] (6.782,3.853)--(7.698,3.853)--(7.698,4.007)--(6.782,4.007)--cycle;
\def\gpfillpath{(3.896,0.985)--(4.169,0.985)--(4.169,1.145)--(3.896,1.145)--cycle}
\gpfill{color=gpbgfillcolor} \gpfillpath;
\gpfill{color=gp lt color 4,gp pattern 4,pattern color=.} \gpfillpath;
\draw[gp path] (3.896,0.985)--(3.896,1.144)--(4.168,1.144)--(4.168,0.985)--cycle;
\def\gpfillpath{(6.066,0.985)--(6.338,0.985)--(6.338,1.569)--(6.066,1.569)--cycle}
\gpfill{color=gpbgfillcolor} \gpfillpath;
\gpfill{color=gp lt color 4,gp pattern 4,pattern color=.} \gpfillpath;
\draw[gp path] (6.066,0.985)--(6.066,1.568)--(6.337,1.568)--(6.337,0.985)--cycle;
\gpcolor{color=gp lt color border}
\node[gp node right] at (6.598,3.622) {max block};
\def\gpfillpath{(6.782,3.545)--(7.698,3.545)--(7.698,3.699)--(6.782,3.699)--cycle}
\gpfill{color=gpbgfillcolor} \gpfillpath;
\gpfill{color=gp lt color 5,gp pattern 5,pattern color=.} \gpfillpath;
\gpcolor{color=gp lt color 5}
\gpsetlinetype{gp lt plot 5}
\draw[gp path] (6.782,3.545)--(7.698,3.545)--(7.698,3.699)--(6.782,3.699)--cycle;
\def\gpfillpath{(4.168,0.985)--(4.440,0.985)--(4.440,1.150)--(4.168,1.150)--cycle}
\gpfill{color=gpbgfillcolor} \gpfillpath;
\gpfill{color=gp lt color 5,gp pattern 5,pattern color=.} \gpfillpath;
\draw[gp path] (4.168,0.985)--(4.168,1.149)--(4.439,1.149)--(4.439,0.985)--cycle;
\def\gpfillpath{(6.337,0.985)--(6.610,0.985)--(6.610,1.510)--(6.337,1.510)--cycle}
\gpfill{color=gpbgfillcolor} \gpfillpath;
\gpfill{color=gp lt color 5,gp pattern 5,pattern color=.} \gpfillpath;
\draw[gp path] (6.337,0.985)--(6.337,1.509)--(6.609,1.509)--(6.609,0.985)--cycle;
\gpcolor{color=gp lt color border}
\gpsetlinetype{gp lt border}
\draw[gp path] (1.320,2.971)--(1.320,0.985)--(7.829,0.985)--(7.829,2.971)--cycle;
\gpdefrectangularnode{gp plot 1}{\pgfpoint{1.320cm}{0.985cm}}{\pgfpoint{7.829cm}{2.971cm}}
\end{tikzpicture}
		\vspace{-15pt}
		\caption{The effectiveness of I/O optimizations on dense matrix
		multiplication.}
		\label{perf:dmm_opts}
	\end{center}
\end{figure}
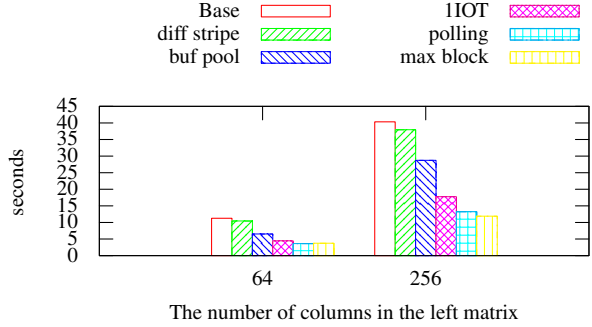

The effectiveness of the optimizations on external memory dense matrix
multiplication is shown in Figure \ref{perf:dmm_opts}. We only demonstrate
their effectiveness on the second form of dense matrix multiplication and
their effectiveness in the first form is very similar. As shown in Figure
\ref{perf:dmm_opts}, using a per-thread buffer pool and a smaller number
of I/O threads has the most significant performance improvement. By combining
all of the optimizations together, we increase the performance by a factor
of up to four. This also indicates the importance of reducing the overhead
of thread context switches in high-throughput I/O access.

\subsubsection{Dense matrix multiplication performance}

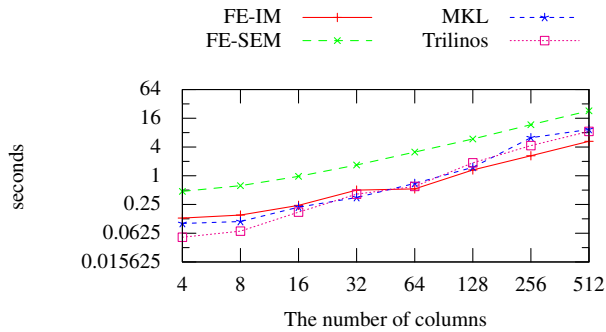
\begin{figure}
	\begin{center}
		\footnotesize
		\vspace{-15pt}
		\begin{tikzpicture}[gnuplot]
\path (0.000,0.000) rectangle (8.382,4.572);
\gpcolor{color=gp lt color border}
\gpsetlinetype{gp lt border}
\gpsetlinewidth{1.00}
\draw[gp path] (2.424,0.985)--(2.604,0.985);
\draw[gp path] (7.829,0.985)--(7.649,0.985);
\node[gp node right] at (2.240,0.985) { 0.015625};
\draw[gp path] (2.424,1.176)--(2.514,1.176);
\draw[gp path] (7.829,1.176)--(7.739,1.176);
\draw[gp path] (2.424,1.367)--(2.604,1.367);
\draw[gp path] (7.829,1.367)--(7.649,1.367);
\node[gp node right] at (2.240,1.367) { 0.0625};
\draw[gp path] (2.424,1.559)--(2.514,1.559);
\draw[gp path] (7.829,1.559)--(7.739,1.559);
\draw[gp path] (2.424,1.750)--(2.604,1.750);
\draw[gp path] (7.829,1.750)--(7.649,1.750);
\node[gp node right] at (2.240,1.750) { 0.25};
\draw[gp path] (2.424,1.941)--(2.514,1.941);
\draw[gp path] (7.829,1.941)--(7.739,1.941);
\draw[gp path] (2.424,2.132)--(2.604,2.132);
\draw[gp path] (7.829,2.132)--(7.649,2.132);
\node[gp node right] at (2.240,2.132) { 1};
\draw[gp path] (2.424,2.323)--(2.514,2.323);
\draw[gp path] (7.829,2.323)--(7.739,2.323);
\draw[gp path] (2.424,2.514)--(2.604,2.514);
\draw[gp path] (7.829,2.514)--(7.649,2.514);
\node[gp node right] at (2.240,2.514) { 4};
\draw[gp path] (2.424,2.706)--(2.514,2.706);
\draw[gp path] (7.829,2.706)--(7.739,2.706);
\draw[gp path] (2.424,2.897)--(2.604,2.897);
\draw[gp path] (7.829,2.897)--(7.649,2.897);
\node[gp node right] at (2.240,2.897) { 16};
\draw[gp path] (2.424,3.088)--(2.514,3.088);
\draw[gp path] (7.829,3.088)--(7.739,3.088);
\draw[gp path] (2.424,3.279)--(2.604,3.279);
\draw[gp path] (7.829,3.279)--(7.649,3.279);
\node[gp node right] at (2.240,3.279) { 64};
\draw[gp path] (2.424,0.985)--(2.424,1.165);
\draw[gp path] (2.424,3.279)--(2.424,3.099);
\node[gp node center] at (2.424,0.677) {4};
\draw[gp path] (3.196,0.985)--(3.196,1.165);
\draw[gp path] (3.196,3.279)--(3.196,3.099);
\node[gp node center] at (3.196,0.677) {8};
\draw[gp path] (3.968,0.985)--(3.968,1.165);
\draw[gp path] (3.968,3.279)--(3.968,3.099);
\node[gp node center] at (3.968,0.677) {16};
\draw[gp path] (4.740,0.985)--(4.740,1.165);
\draw[gp path] (4.740,3.279)--(4.740,3.099);
\node[gp node center] at (4.740,0.677) {32};
\draw[gp path] (5.513,0.985)--(5.513,1.165);
\draw[gp path] (5.513,3.279)--(5.513,3.099);
\node[gp node center] at (5.513,0.677) {64};
\draw[gp path] (6.285,0.985)--(6.285,1.165);
\draw[gp path] (6.285,3.279)--(6.285,3.099);
\node[gp node center] at (6.285,0.677) {128};
\draw[gp path] (7.057,0.985)--(7.057,1.165);
\draw[gp path] (7.057,3.279)--(7.057,3.099);
\node[gp node center] at (7.057,0.677) {256};
\draw[gp path] (7.829,0.985)--(7.829,1.165);
\draw[gp path] (7.829,3.279)--(7.829,3.099);
\node[gp node center] at (7.829,0.677) {512};
\draw[gp path] (2.424,3.279)--(2.424,0.985)--(7.829,0.985)--(7.829,3.279)--cycle;
\node[gp node center,rotate=-270] at (0.246,2.132) {seconds};
\node[gp node center] at (5.126,0.215) {The number of columns};
\node[gp node right] at (3.842,4.238) {FE-IM};
\gpcolor{color=gp lt color 0}
\gpsetlinetype{gp lt plot 0}
\draw[gp path] (4.026,4.238)--(4.942,4.238);
\draw[gp path] (2.424,1.569)--(3.196,1.609)--(3.968,1.738)--(4.740,1.941)--(5.513,1.957)%
  --(6.285,2.209)--(7.057,2.398)--(7.829,2.590);
\gpsetpointsize{4.00}
\gppoint{gp mark 1}{(2.424,1.569)}
\gppoint{gp mark 1}{(3.196,1.609)}
\gppoint{gp mark 1}{(3.968,1.738)}
\gppoint{gp mark 1}{(4.740,1.941)}
\gppoint{gp mark 1}{(5.513,1.957)}
\gppoint{gp mark 1}{(6.285,2.209)}
\gppoint{gp mark 1}{(7.057,2.398)}
\gppoint{gp mark 1}{(7.829,2.590)}
\gppoint{gp mark 1}{(4.484,4.238)}
\gpcolor{color=gp lt color border}
\node[gp node right] at (3.842,3.930) {FE-SEM};
\gpcolor{color=gp lt color 1}
\gpsetlinetype{gp lt plot 1}
\draw[gp path] (4.026,3.930)--(4.942,3.930);
\draw[gp path] (2.424,1.924)--(3.196,2.000)--(3.968,2.124)--(4.740,2.275)--(5.513,2.447)%
  --(6.285,2.621)--(7.057,2.811)--(7.829,2.996);
\gppoint{gp mark 2}{(2.424,1.924)}
\gppoint{gp mark 2}{(3.196,2.000)}
\gppoint{gp mark 2}{(3.968,2.124)}
\gppoint{gp mark 2}{(4.740,2.275)}
\gppoint{gp mark 2}{(5.513,2.447)}
\gppoint{gp mark 2}{(6.285,2.621)}
\gppoint{gp mark 2}{(7.057,2.811)}
\gppoint{gp mark 2}{(7.829,2.996)}
\gppoint{gp mark 2}{(4.484,3.930)}
\gpcolor{color=gp lt color border}
\node[gp node right] at (6.598,4.238) {MKL};
\gpcolor{color=gp lt color 2}
\gpsetlinetype{gp lt plot 2}
\draw[gp path] (6.782,4.238)--(7.698,4.238);
\draw[gp path] (2.424,1.497)--(3.196,1.523)--(3.968,1.714)--(4.740,1.842)--(5.513,2.030)%
  --(6.285,2.244)--(7.057,2.638)--(7.829,2.748);
\gppoint{gp mark 3}{(2.424,1.497)}
\gppoint{gp mark 3}{(3.196,1.523)}
\gppoint{gp mark 3}{(3.968,1.714)}
\gppoint{gp mark 3}{(4.740,1.842)}
\gppoint{gp mark 3}{(5.513,2.030)}
\gppoint{gp mark 3}{(6.285,2.244)}
\gppoint{gp mark 3}{(7.057,2.638)}
\gppoint{gp mark 3}{(7.829,2.748)}
\gppoint{gp mark 3}{(7.240,4.238)}
\gpcolor{color=gp lt color border}
\node[gp node right] at (6.598,3.930) {Trilinos};
\gpcolor{color=gp lt color 3}
\gpsetlinetype{gp lt plot 3}
\draw[gp path] (6.782,3.930)--(7.698,3.930);
\draw[gp path] (2.424,1.314)--(3.196,1.395)--(3.968,1.648)--(4.740,1.887)--(5.513,1.990)%
  --(6.285,2.304)--(7.057,2.533)--(7.829,2.722);
\gppoint{gp mark 4}{(2.424,1.314)}
\gppoint{gp mark 4}{(3.196,1.395)}
\gppoint{gp mark 4}{(3.968,1.648)}
\gppoint{gp mark 4}{(4.740,1.887)}
\gppoint{gp mark 4}{(5.513,1.990)}
\gppoint{gp mark 4}{(6.285,2.304)}
\gppoint{gp mark 4}{(7.057,2.533)}
\gppoint{gp mark 4}{(7.829,2.722)}
\gppoint{gp mark 4}{(7.240,3.930)}
\gpcolor{color=gp lt color border}
\gpsetlinetype{gp lt border}
\draw[gp path] (2.424,3.279)--(2.424,0.985)--(7.829,0.985)--(7.829,3.279)--cycle;
\gpdefrectangularnode{gp plot 1}{\pgfpoint{2.424cm}{0.985cm}}{\pgfpoint{7.829cm}{3.279cm}}
\end{tikzpicture}
		\vspace{-15pt}
		\caption{The runtime of dense matrix multiplication in $op1$. We compare
			in-memory (FE-IM) and external-memory (FE-EM) implementations
			in FlashEigen with the MKL and Trilinos implementations.}
		\label{perf:gemm}
	\end{center}
\end{figure}

Figure \ref{perf:gemm} show the performance of the in-memory and external-memory
dense matrix multiplication in the first form. We omit the performance result
of dense matrix multiplication in the second form because it is very similar
to the first form and MKL cannot parallelize it.
The external-memory multiplication in FlashEigen is roughly 3-6 times slower
than its in-memory counterpart. In the dense matrix multiplication of
the first form, the in-memory implementation outperforms MKL and Trilinos
when the number of the columns get larger and MKL does not have a parallel
implementation for the dense matrix multiplication of the second form.

\begin{figure}
	\begin{center}
		\footnotesize
		\vspace{-15pt}
		\begin{tikzpicture}[gnuplot]
\path (0.000,0.000) rectangle (8.382,4.572);
\gpcolor{color=gp lt color border}
\gpsetlinetype{gp lt border}
\gpsetlinewidth{1.00}
\draw[gp path] (1.320,0.985)--(1.500,0.985);
\draw[gp path] (7.829,0.985)--(7.649,0.985);
\node[gp node right] at (1.136,0.985) { 0};
\draw[gp path] (1.320,1.419)--(1.500,1.419);
\draw[gp path] (7.829,1.419)--(7.649,1.419);
\node[gp node right] at (1.136,1.419) { 2};
\draw[gp path] (1.320,1.852)--(1.500,1.852);
\draw[gp path] (7.829,1.852)--(7.649,1.852);
\node[gp node right] at (1.136,1.852) { 4};
\draw[gp path] (1.320,2.286)--(1.500,2.286);
\draw[gp path] (7.829,2.286)--(7.649,2.286);
\node[gp node right] at (1.136,2.286) { 6};
\draw[gp path] (1.320,2.720)--(1.500,2.720);
\draw[gp path] (7.829,2.720)--(7.649,2.720);
\node[gp node right] at (1.136,2.720) { 8};
\draw[gp path] (1.320,3.153)--(1.500,3.153);
\draw[gp path] (7.829,3.153)--(7.649,3.153);
\node[gp node right] at (1.136,3.153) { 10};
\draw[gp path] (1.320,3.587)--(1.500,3.587);
\draw[gp path] (7.829,3.587)--(7.649,3.587);
\node[gp node right] at (1.136,3.587) { 12};
\draw[gp path] (1.320,0.985)--(1.320,1.165);
\draw[gp path] (1.320,3.587)--(1.320,3.407);
\node[gp node center] at (1.320,0.677) {4};
\draw[gp path] (2.250,0.985)--(2.250,1.165);
\draw[gp path] (2.250,3.587)--(2.250,3.407);
\node[gp node center] at (2.250,0.677) {8};
\draw[gp path] (3.180,0.985)--(3.180,1.165);
\draw[gp path] (3.180,3.587)--(3.180,3.407);
\node[gp node center] at (3.180,0.677) {16};
\draw[gp path] (4.110,0.985)--(4.110,1.165);
\draw[gp path] (4.110,3.587)--(4.110,3.407);
\node[gp node center] at (4.110,0.677) {32};
\draw[gp path] (5.039,0.985)--(5.039,1.165);
\draw[gp path] (5.039,3.587)--(5.039,3.407);
\node[gp node center] at (5.039,0.677) {64};
\draw[gp path] (5.969,0.985)--(5.969,1.165);
\draw[gp path] (5.969,3.587)--(5.969,3.407);
\node[gp node center] at (5.969,0.677) {128};
\draw[gp path] (6.899,0.985)--(6.899,1.165);
\draw[gp path] (6.899,3.587)--(6.899,3.407);
\node[gp node center] at (6.899,0.677) {256};
\draw[gp path] (7.829,0.985)--(7.829,1.165);
\draw[gp path] (7.829,3.587)--(7.829,3.407);
\node[gp node center] at (7.829,0.677) {512};
\draw[gp path] (1.320,3.587)--(1.320,0.985)--(7.829,0.985)--(7.829,3.587)--cycle;
\node[gp node center,rotate=-270] at (0.246,2.286) {GB/s};
\node[gp node center] at (4.574,0.215) {The number of columns};
\node[gp node right] at (3.290,4.238) {op1};
\gpcolor{color=gp lt color 0}
\gpsetlinetype{gp lt plot 0}
\draw[gp path] (3.474,4.238)--(4.390,4.238);
\draw[gp path] (1.320,2.724)--(2.250,2.967)--(3.180,3.075)--(4.110,3.162)--(5.039,3.197)%
  --(5.969,3.266)--(6.899,3.242)--(7.829,3.270);
\gpsetpointsize{4.00}
\gppoint{gp mark 1}{(1.320,2.724)}
\gppoint{gp mark 1}{(2.250,2.967)}
\gppoint{gp mark 1}{(3.180,3.075)}
\gppoint{gp mark 1}{(4.110,3.162)}
\gppoint{gp mark 1}{(5.039,3.197)}
\gppoint{gp mark 1}{(5.969,3.266)}
\gppoint{gp mark 1}{(6.899,3.242)}
\gppoint{gp mark 1}{(7.829,3.270)}
\gppoint{gp mark 1}{(3.932,4.238)}
\gpcolor{color=gp lt color border}
\node[gp node right] at (5.126,4.238) {op3};
\gpcolor{color=gp lt color 1}
\gpsetlinetype{gp lt plot 1}
\draw[gp path] (5.310,4.238)--(6.226,4.238);
\draw[gp path] (1.320,2.785)--(2.250,3.088)--(3.180,3.129)--(4.110,3.197)--(5.039,3.225)%
  --(5.969,3.312)--(6.899,3.290)--(7.829,3.342);
\gppoint{gp mark 2}{(1.320,2.785)}
\gppoint{gp mark 2}{(2.250,3.088)}
\gppoint{gp mark 2}{(3.180,3.129)}
\gppoint{gp mark 2}{(4.110,3.197)}
\gppoint{gp mark 2}{(5.039,3.225)}
\gppoint{gp mark 2}{(5.969,3.312)}
\gppoint{gp mark 2}{(6.899,3.290)}
\gppoint{gp mark 2}{(7.829,3.342)}
\gppoint{gp mark 2}{(5.768,4.238)}
\gpcolor{color=gp lt color border}
\gpsetlinetype{gp lt border}
\draw[gp path] (1.320,3.587)--(1.320,0.985)--(7.829,0.985)--(7.829,3.587)--cycle;
\gpdefrectangularnode{gp plot 1}{\pgfpoint{1.320cm}{0.985cm}}{\pgfpoint{7.829cm}{3.587cm}}
\end{tikzpicture}
		\vspace{-15pt}
		\caption{The average I/O throughput in dense matrix multiplication
		on the SSD array.}
		\label{perf:dmm_io}
	\end{center}
\end{figure}
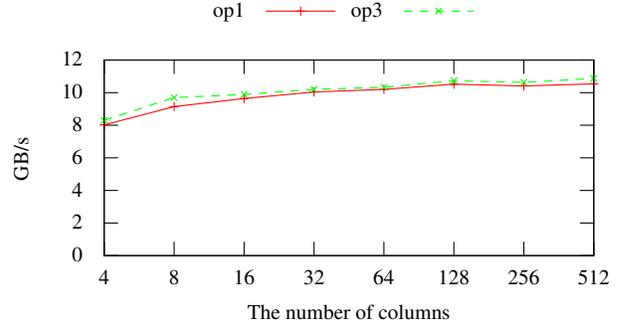

The external-memory dense matrix multiplication has almost saturated
the I/O bandwidth of the SSDs (Figure \ref{perf:dmm_io}). The average
I/O throughput reaches 10.87GB/s from the entire SSD array or 464MB/s
per SSD, while the peak throughput of an SSD is approaching 500MB/s,
which is close to the maximal I/O throughput of 540MB/s per SSD, advertised by
the SSD vendor. This also indicates that the SSDs are the bottleneck for
the dense matrix multiplication in the eigensolver. This is not
surprising because the sequential I/O performance of SSDs is an order of
magnitude smaller than RAM. 

\subsection{The KrylovSchur eigensolver in FlashEigen}
In this section, we focus on evaluating the performance of the KrylovSchur
eigensolvers in FlashEigen. KrylovSchur is not only the fastest in-memory
eigensolver on the graphs in Table \ref{graphs} among all Anasazi eigensolvers,
but also generates the least I/O, especially writes,
to SSDs. Reducing I/O is essential to achieve performance and reduce SSD
wearout. We evaluate our external-memory eigensolver and compare it
with its in-memory counterpart as well as the original KrylovSchur eigensolver
in the Anasazi framework.

The KrylovSchur eigensolver has two important parameters: the subspace size
and the block size. They significantly affect the convergence of
the KrylovSchur eigensolver. A reasonably large subspace size and block size
accelerates its convergence. However, a larger subspace increases memory
consumption of the eigensolver as well as computation and I/O complexity
in a single matrix operation. A larger block size increases memory consumption
of FlashEigen.

For the experiments below, we select the values for the two parameters that
achieve the best runtime performance for the eigensolvers. We assume that
our setting is not constrained by memory size available to the machine.
Because sparse matrix in Trilinos is not optimized for
the dense matrix with more than one column, we use $1$ as the block size and
$2 \times ev$ as the number of blocks for most of the graphs in the original
KrylovSchur eigensolver, where $ev$ is the number of the eigenvalues to compute.
For the FlashEigen KrylovSchur eigensolver, we choose the subspace size and
the block size to reduce the amount of I/O to SSDs. we use $1$ as the block size and
$2 \times ev$ as the number of blocks when computing a small number of eigenvalues
and use $4$ as the block size and $ev$ as the number of blocks. Therefore,
the FlashEigen eigensolver uses the subspace twice as large as the one used by
the original KrylovSchur eigensolver when computing a large number of eigenvalues.
However, the eigenvalues of the graph W are very close to each other, so we have
to use a much larger subspace to have the eigensolver to converge or converge
faster. For the W graph, we use the subspace four times larger than the ones
used for other graphs. Therefore, we can trade off the runtime performance
with memory consumption in eigendecomposition.

\subsubsection{Performance of eigensolvers}

We evaluate the performance of our SEM eigensolver and compare its performance
with our in-memory eigensolver and the original KrylovSchur eigensolver.
We compute different numbers of eigenvalues on the three smaller graphs in
Table \ref{graphs}. Only our SEM eigensolver is able to compute eigenvalues
on the page graph on the 1TB-memory machine.

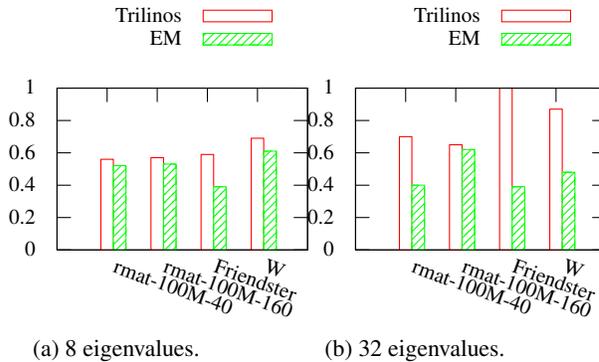
\begin{figure}[t]
\centering
\footnotesize
\vspace{-15pt}
\begin{subfigure}{.5\linewidth}
	\begin{tikzpicture}[gnuplot]
\path (0.000,0.000) rectangle (5.080,4.572);
\gpcolor{color=gp lt color border}
\gpsetlinetype{gp lt border}
\gpsetlinewidth{1.00}
\draw[gp path] (1.196,1.126)--(1.376,1.126);
\draw[gp path] (4.527,1.126)--(4.347,1.126);
\node[gp node right] at (1.012,1.126) { 0};
\draw[gp path] (1.196,1.557)--(1.376,1.557);
\draw[gp path] (4.527,1.557)--(4.347,1.557);
\node[gp node right] at (1.012,1.557) { 0.2};
\draw[gp path] (1.196,1.987)--(1.376,1.987);
\draw[gp path] (4.527,1.987)--(4.347,1.987);
\node[gp node right] at (1.012,1.987) { 0.4};
\draw[gp path] (1.196,2.418)--(1.376,2.418);
\draw[gp path] (4.527,2.418)--(4.347,2.418);
\node[gp node right] at (1.012,2.418) { 0.6};
\draw[gp path] (1.196,2.848)--(1.376,2.848);
\draw[gp path] (4.527,2.848)--(4.347,2.848);
\node[gp node right] at (1.012,2.848) { 0.8};
\draw[gp path] (1.196,3.279)--(1.376,3.279);
\draw[gp path] (4.527,3.279)--(4.347,3.279);
\node[gp node right] at (1.012,3.279) { 1};
\draw[gp path] (1.862,1.126)--(1.862,1.306);
\draw[gp path] (1.862,3.279)--(1.862,3.099);
\node[gp node left,rotate=-20] at (1.862,0.942) {rmat-100M-40};
\draw[gp path] (2.528,1.126)--(2.528,1.306);
\draw[gp path] (2.528,3.279)--(2.528,3.099);
\node[gp node left,rotate=-20] at (2.528,0.942) {rmat-100M-160};
\draw[gp path] (3.195,1.126)--(3.195,1.306);
\draw[gp path] (3.195,3.279)--(3.195,3.099);
\node[gp node left,rotate=-20] at (3.195,0.942) {Friendster};
\draw[gp path] (3.861,1.126)--(3.861,1.306);
\draw[gp path] (3.861,3.279)--(3.861,3.099);
\node[gp node left,rotate=-20] at (3.861,0.942) {W};
\draw[gp path] (1.196,3.279)--(1.196,1.126)--(4.527,1.126)--(4.527,3.279)--cycle;
\node[gp node right] at (2.955,4.238) {Trilinos};
\def\gpfillpath{(3.139,4.161)--(4.055,4.161)--(4.055,4.315)--(3.139,4.315)--cycle}
\gpfill{color=gpbgfillcolor} \gpfillpath;
\gpfill{color=gp lt color 0,gp pattern 0,pattern color=.} \gpfillpath;
\gpcolor{color=gp lt color 0}
\gpsetlinetype{gp lt plot 0}
\draw[gp path] (3.139,4.161)--(4.055,4.161)--(4.055,4.315)--(3.139,4.315)--cycle;
\def\gpfillpath{(1.779,1.126)--(1.946,1.126)--(1.946,2.333)--(1.779,2.333)--cycle}
\gpfill{color=gpbgfillcolor} \gpfillpath;
\gpfill{color=gp lt color 0,gp pattern 0,pattern color=.} \gpfillpath;
\draw[gp path] (1.779,1.126)--(1.779,2.332)--(1.945,2.332)--(1.945,1.126)--cycle;
\def\gpfillpath{(2.445,1.126)--(2.613,1.126)--(2.613,2.354)--(2.445,2.354)--cycle}
\gpfill{color=gpbgfillcolor} \gpfillpath;
\gpfill{color=gp lt color 0,gp pattern 0,pattern color=.} \gpfillpath;
\draw[gp path] (2.445,1.126)--(2.445,2.353)--(2.612,2.353)--(2.612,1.126)--cycle;
\def\gpfillpath{(3.111,1.126)--(3.279,1.126)--(3.279,2.397)--(3.111,2.397)--cycle}
\gpfill{color=gpbgfillcolor} \gpfillpath;
\gpfill{color=gp lt color 0,gp pattern 0,pattern color=.} \gpfillpath;
\draw[gp path] (3.111,1.126)--(3.111,2.396)--(3.278,2.396)--(3.278,1.126)--cycle;
\def\gpfillpath{(3.778,1.126)--(3.945,1.126)--(3.945,2.613)--(3.778,2.613)--cycle}
\gpfill{color=gpbgfillcolor} \gpfillpath;
\gpfill{color=gp lt color 0,gp pattern 0,pattern color=.} \gpfillpath;
\draw[gp path] (3.778,1.126)--(3.778,2.612)--(3.944,2.612)--(3.944,1.126)--cycle;
\gpcolor{color=gp lt color border}
\node[gp node right] at (2.955,3.930) {EM};
\def\gpfillpath{(3.139,3.853)--(4.055,3.853)--(4.055,4.007)--(3.139,4.007)--cycle}
\gpfill{color=gpbgfillcolor} \gpfillpath;
\gpfill{color=gp lt color 1,gp pattern 1,pattern color=.} \gpfillpath;
\gpcolor{color=gp lt color 1}
\gpsetlinetype{gp lt plot 1}
\draw[gp path] (3.139,3.853)--(4.055,3.853)--(4.055,4.007)--(3.139,4.007)--cycle;
\def\gpfillpath{(1.945,1.126)--(2.113,1.126)--(2.113,2.247)--(1.945,2.247)--cycle}
\gpfill{color=gpbgfillcolor} \gpfillpath;
\gpfill{color=gp lt color 1,gp pattern 1,pattern color=.} \gpfillpath;
\draw[gp path] (1.945,1.126)--(1.945,2.246)--(2.112,2.246)--(2.112,1.126)--cycle;
\def\gpfillpath{(2.612,1.126)--(2.779,1.126)--(2.779,2.268)--(2.612,2.268)--cycle}
\gpfill{color=gpbgfillcolor} \gpfillpath;
\gpfill{color=gp lt color 1,gp pattern 1,pattern color=.} \gpfillpath;
\draw[gp path] (2.612,1.126)--(2.612,2.267)--(2.778,2.267)--(2.778,1.126)--cycle;
\def\gpfillpath{(3.278,1.126)--(3.445,1.126)--(3.445,1.967)--(3.278,1.967)--cycle}
\gpfill{color=gpbgfillcolor} \gpfillpath;
\gpfill{color=gp lt color 1,gp pattern 1,pattern color=.} \gpfillpath;
\draw[gp path] (3.278,1.126)--(3.278,1.966)--(3.444,1.966)--(3.444,1.126)--cycle;
\def\gpfillpath{(3.944,1.126)--(4.112,1.126)--(4.112,2.440)--(3.944,2.440)--cycle}
\gpfill{color=gpbgfillcolor} \gpfillpath;
\gpfill{color=gp lt color 1,gp pattern 1,pattern color=.} \gpfillpath;
\draw[gp path] (3.944,1.126)--(3.944,2.439)--(4.111,2.439)--(4.111,1.126)--cycle;
\gpcolor{color=gp lt color border}
\gpsetlinetype{gp lt border}
\draw[gp path] (1.196,3.279)--(1.196,1.126)--(4.527,1.126)--(4.527,3.279)--cycle;
\gpdefrectangularnode{gp plot 1}{\pgfpoint{1.196cm}{1.126cm}}{\pgfpoint{4.527cm}{3.279cm}}
\end{tikzpicture}
	\vspace{-15pt}
	\caption{8 eigenvalues.}
	\label{fig:eigen8}
\end{subfigure}%
\begin{subfigure}{.5\linewidth}
	\begin{tikzpicture}[gnuplot]
\path (0.000,0.000) rectangle (5.080,4.572);
\gpcolor{color=gp lt color border}
\gpsetlinetype{gp lt border}
\gpsetlinewidth{1.00}
\draw[gp path] (1.196,1.126)--(1.376,1.126);
\draw[gp path] (4.527,1.126)--(4.347,1.126);
\node[gp node right] at (1.012,1.126) { 0};
\draw[gp path] (1.196,1.557)--(1.376,1.557);
\draw[gp path] (4.527,1.557)--(4.347,1.557);
\node[gp node right] at (1.012,1.557) { 0.2};
\draw[gp path] (1.196,1.987)--(1.376,1.987);
\draw[gp path] (4.527,1.987)--(4.347,1.987);
\node[gp node right] at (1.012,1.987) { 0.4};
\draw[gp path] (1.196,2.418)--(1.376,2.418);
\draw[gp path] (4.527,2.418)--(4.347,2.418);
\node[gp node right] at (1.012,2.418) { 0.6};
\draw[gp path] (1.196,2.848)--(1.376,2.848);
\draw[gp path] (4.527,2.848)--(4.347,2.848);
\node[gp node right] at (1.012,2.848) { 0.8};
\draw[gp path] (1.196,3.279)--(1.376,3.279);
\draw[gp path] (4.527,3.279)--(4.347,3.279);
\node[gp node right] at (1.012,3.279) { 1};
\draw[gp path] (1.862,1.126)--(1.862,1.306);
\draw[gp path] (1.862,3.279)--(1.862,3.099);
\node[gp node left,rotate=-20] at (1.862,0.942) {rmat-100M-40};
\draw[gp path] (2.528,1.126)--(2.528,1.306);
\draw[gp path] (2.528,3.279)--(2.528,3.099);
\node[gp node left,rotate=-20] at (2.528,0.942) {rmat-100M-160};
\draw[gp path] (3.195,1.126)--(3.195,1.306);
\draw[gp path] (3.195,3.279)--(3.195,3.099);
\node[gp node left,rotate=-20] at (3.195,0.942) {Friendster};
\draw[gp path] (3.861,1.126)--(3.861,1.306);
\draw[gp path] (3.861,3.279)--(3.861,3.099);
\node[gp node left,rotate=-20] at (3.861,0.942) {W};
\draw[gp path] (1.196,3.279)--(1.196,1.126)--(4.527,1.126)--(4.527,3.279)--cycle;
\node[gp node right] at (2.955,4.238) {Trilinos};
\def\gpfillpath{(3.139,4.161)--(4.055,4.161)--(4.055,4.315)--(3.139,4.315)--cycle}
\gpfill{color=gpbgfillcolor} \gpfillpath;
\gpfill{color=gp lt color 0,gp pattern 0,pattern color=.} \gpfillpath;
\gpcolor{color=gp lt color 0}
\gpsetlinetype{gp lt plot 0}
\draw[gp path] (3.139,4.161)--(4.055,4.161)--(4.055,4.315)--(3.139,4.315)--cycle;
\def\gpfillpath{(1.779,1.126)--(1.946,1.126)--(1.946,2.634)--(1.779,2.634)--cycle}
\gpfill{color=gpbgfillcolor} \gpfillpath;
\gpfill{color=gp lt color 0,gp pattern 0,pattern color=.} \gpfillpath;
\draw[gp path] (1.779,1.126)--(1.779,2.633)--(1.945,2.633)--(1.945,1.126)--cycle;
\def\gpfillpath{(2.445,1.126)--(2.613,1.126)--(2.613,2.526)--(2.445,2.526)--cycle}
\gpfill{color=gpbgfillcolor} \gpfillpath;
\gpfill{color=gp lt color 0,gp pattern 0,pattern color=.} \gpfillpath;
\draw[gp path] (2.445,1.126)--(2.445,2.525)--(2.612,2.525)--(2.612,1.126)--cycle;
\def\gpfillpath{(3.111,1.126)--(3.279,1.126)--(3.279,3.280)--(3.111,3.280)--cycle}
\gpfill{color=gpbgfillcolor} \gpfillpath;
\gpfill{color=gp lt color 0,gp pattern 0,pattern color=.} \gpfillpath;
\draw[gp path] (3.111,1.126)--(3.111,3.279)--(3.278,3.279)--(3.278,1.126)--cycle;
\def\gpfillpath{(3.778,1.126)--(3.945,1.126)--(3.945,3.000)--(3.778,3.000)--cycle}
\gpfill{color=gpbgfillcolor} \gpfillpath;
\gpfill{color=gp lt color 0,gp pattern 0,pattern color=.} \gpfillpath;
\draw[gp path] (3.778,1.126)--(3.778,2.999)--(3.944,2.999)--(3.944,1.126)--cycle;
\gpcolor{color=gp lt color border}
\node[gp node right] at (2.955,3.930) {EM};
\def\gpfillpath{(3.139,3.853)--(4.055,3.853)--(4.055,4.007)--(3.139,4.007)--cycle}
\gpfill{color=gpbgfillcolor} \gpfillpath;
\gpfill{color=gp lt color 1,gp pattern 1,pattern color=.} \gpfillpath;
\gpcolor{color=gp lt color 1}
\gpsetlinetype{gp lt plot 1}
\draw[gp path] (3.139,3.853)--(4.055,3.853)--(4.055,4.007)--(3.139,4.007)--cycle;
\def\gpfillpath{(1.945,1.126)--(2.113,1.126)--(2.113,1.988)--(1.945,1.988)--cycle}
\gpfill{color=gpbgfillcolor} \gpfillpath;
\gpfill{color=gp lt color 1,gp pattern 1,pattern color=.} \gpfillpath;
\draw[gp path] (1.945,1.126)--(1.945,1.987)--(2.112,1.987)--(2.112,1.126)--cycle;
\def\gpfillpath{(2.612,1.126)--(2.779,1.126)--(2.779,2.462)--(2.612,2.462)--cycle}
\gpfill{color=gpbgfillcolor} \gpfillpath;
\gpfill{color=gp lt color 1,gp pattern 1,pattern color=.} \gpfillpath;
\draw[gp path] (2.612,1.126)--(2.612,2.461)--(2.778,2.461)--(2.778,1.126)--cycle;
\def\gpfillpath{(3.278,1.126)--(3.445,1.126)--(3.445,1.967)--(3.278,1.967)--cycle}
\gpfill{color=gpbgfillcolor} \gpfillpath;
\gpfill{color=gp lt color 1,gp pattern 1,pattern color=.} \gpfillpath;
\draw[gp path] (3.278,1.126)--(3.278,1.966)--(3.444,1.966)--(3.444,1.126)--cycle;
\def\gpfillpath{(3.944,1.126)--(4.112,1.126)--(4.112,2.160)--(3.944,2.160)--cycle}
\gpfill{color=gpbgfillcolor} \gpfillpath;
\gpfill{color=gp lt color 1,gp pattern 1,pattern color=.} \gpfillpath;
\draw[gp path] (3.944,1.126)--(3.944,2.159)--(4.111,2.159)--(4.111,1.126)--cycle;
\gpcolor{color=gp lt color border}
\gpsetlinetype{gp lt border}
\draw[gp path] (1.196,3.279)--(1.196,1.126)--(4.527,1.126)--(4.527,3.279)--cycle;
\gpdefrectangularnode{gp plot 1}{\pgfpoint{1.196cm}{1.126cm}}{\pgfpoint{4.527cm}{3.279cm}}
\end{tikzpicture}
	\vspace{-15pt}
	\caption{32 eigenvalues.}
	\label{fig:eigen32}
\end{subfigure}
\caption{The performance of the Trilinos KrylovSchur and FlashEigen-EM
KrylovSchur relative to the FlashEigen-IM KrylovSchur.}
\vspace{-15pt}
\label{fig:eigen}
\end{figure}

Our SEM eigensolver achieves at least 40\% performance of our in-memory
eigensolver, while the in-memory eigensolver outperforms the original
KrylovSchur eigensolver (Figure \ref{fig:eigen}). The SEM eigensolver
is more efficient to compute a small number of eigenvalues and is able
to achieve around 50\% performance of our in-memory eigensolver.
SpMM and reorthogonalization
account for most of computation time when computing a small number of eigenvalues,
but reorthogonalization eventually dominates the eigendecomposition for computing
many eigenvalues. Because external-memory dense matrix multiplication is several
times slower than the in-memory implementations, reorthogonalization accounts for
over 90\% of runtime in the SEM eigensolver for computing a large number of
eigenvalues. For many spectral analysis tasks, users only require a small number
of eigenvalues.

The SEM eigensolver uses a small fraction of memory used by its in-memory
counterparts and the original Trilinos eigensolver and its memory consumption
remains roughly the same as the number of eigenvalues computed by the eigensolvers
increases. A small memory consumption provides two benefits. First, FlashEigen
is able to scale to a much larger eigenvalue problem. The second benefit is that
FlashEigen gives users more freedom to choose the subspace size that gives the
fastest convergence in a given eigenvalue problem because SSDs significantly
increases memory available to the eigensolver. We have to point out that our
experiments assume that we choose the subspace size that is not constrained by
the memory size in a machine.


\subsubsection{Scale to billion-node graphs}

We evaluate the scalability of FlashEigen with the page graph with 3.4 billion
vertices and 129 billion edges. Because the page graph is a directed graph,
its adjacency matrix is asymmetric and we perform singular value decomposition
(SVD) on the adjacency matrix instead of simple eigendecomposition. For the page
graph, we use $2$ for the block size and $2 \times ev$ for the number of blocks
because sparse matrix multiplication is completely bottlenecked by SSDs.
Neither the in-memory eigensolver nor the original Trilinos eigensolver is able
to compute eigenvalues on the page graph with 1TB RAM.

\begin{table}
	\begin{center}
		\small
		\begin{tabular}{|c|c|c|c|c|}
			\hline
			\#eigenvalues & runtime & memory & read & write \\
			\hline
			8 & 4.2 hours & 120GB & 145TB & 4TB \\
			\hline
		\end{tabular}
		\normalsize
	\end{center}
	\caption{The performance and resource consumption of computing eigenvalues
	on the page graph.}
	\label{pg_ev}
\end{table}

FlashEigen computes a fairly large number of eigenvalues within a reasonable
amount of time and consumes a fairly small amount of resources given the large
size of the eigenvalue problem.
The average I/O throughput during the computation of 8 eigenvalues is about
10GB/s, which is very close to the maximal I/O throughput provided by
the SSD array.
Given the relative small memory footprint, we are able to scale FlashEigen
to a much larger eigenvalue problem on our 1TB-RAM machine.

\section{Conclusions}
We present an external-memory framework called FlashEigen using a large array
of commodity SSDs to solve eigenvalue problems at the billion scale. FlashEigen
utilizes the Anasazi framework to get state-of-art eigensolver implementations
so that we can focus on optimizations on sparse matrix multiplication and dense
matrix multiplication on SSDs.

We implement a sparse matrix dense matrix multiplication in the semi-external
memory fashion. That is, we keep the sparse matrix on SSDs and the dense matrices
in memory. We deploy a set of memory and I/O optimizations so that the sparse
matrix dense matrix multiplication has performance comparable to its in-memory
counterparts while significantly outperforming the MKL and Trilinos implementation.
The semi-external sparse matrix multiplication is able to saturate either CPU or
SSDs or both, which suggests that we have achieved the maximal performance from
the existing hardware.

We further implement and optimize external-memory matrix multiplication for
FlashEigen. When computing eigenvalues on many real-world graphs, the storage
size required by the subspace in an eigensolver is massive. Therefore, we keep
the subspace on the SSD array and deploy a set of I/O optimizations on dense
matrix multiplication. With the I/O optimizations, we saturate the SSDs in
dense matrix multiplication and achieve the maximal performance from the existing
hardware. However, the SSD array is an order of magnitude slower than RAM, so
external-memory dense matrix multiplication is about three or four times slower
than the state-of-art in-memory dense matrix multiplication.

We implement our external-memory eigensolver with the semi-external memory
sparse matrix multiplication and the external-memory dense matrix multiplication.
Our experiments show that our external-memory eigensolver achieves at least
30\% of the performance of its in-memory counterparts. For a small number of
eigenvalues, which is the most common case for spectral analysis, our external-memory
eigensolver has less than 50\% performance loss. We further show that our external-
memory eigensolver is able to scale a graph with 3.4 billion vertices and 129
billion edges. It finishes eigendecomposition within a reasonable amount of time
and consumes a fairly small amount of resources. This suggests that our external-
memory eigensolver is able to scale to even a larger eigenvalue problem in our
1TB-RAM machine.

We are able to scale to much larger graphs.
Given the same amount of resources, SSDs significantly extends the memory capacity
and give users the freedom of choosing the optimal subspace size for better convergence.
Our solution also works better for relatively denser graphs.
Our solution works better for computing a small number of eigenvalues.

\section{Acknowledgments}
This work is partially supported by NSF ACI-1261715,
DARPA GRAPHS N66001-14-1-4028 and
DARPA SIMPLEX program through SPAWAR contract N66001-15-C-4041.

{\footnotesize \bibliographystyle{acm}
\bibliography{eigensolver}  


\end{document}